\documentclass[preprint,superscriptaddress,showpacs,tightenlines]{revtex4}
\usepackage{amssymb}
\usepackage{amsmath}
\usepackage{graphicx,bm}

\input{tcilatex}

\begin{document}

\title{Signature of Fermi surface anisotropy in point contact
conductance in the presence of defects }
\author{Ye.S. Avotina}
\affiliation{B.I. Verkin Institute for Low Temperature Physics and Engineering, National
Academy of Sciences of Ukraine, 47, Lenin Ave., 61103, Kharkov,Ukraine.}
\affiliation{Kamerlingh Onnes Laboratorium, Universiteit Leiden, Postbus 9504, 2300
Leiden, The Netherlands.}
\author{Yu.A. Kolesnichenko}
\affiliation{B.I. Verkin Institute for Low Temperature Physics and Engineering, National
Academy of Sciences of Ukraine, 47, Lenin Ave., 61103, Kharkov,Ukraine.}
\affiliation{Kamerlingh Onnes Laboratorium, Universiteit Leiden, Postbus 9504, 2300
Leiden, The Netherlands.}
\author{A.F. Otte}
\affiliation{Kamerlingh Onnes Laboratorium, Universiteit Leiden, Postbus 9504, 2300
Leiden, The Netherlands.}
\author{J.M. van Ruitenbeek}
\affiliation{Kamerlingh Onnes Laboratorium, Universiteit Leiden, Postbus 9504, 2300
Leiden, The Netherlands.}

\begin{abstract}
In a previous paper (Avotina \textit{et al.}\ Phys. Rev. B
\textbf{71}, 115430 (2005)) we have shown that in principle it is
possible to image the defect positions below a metal surface by
means of a scanning tunnelling microscope. The principle relies on
the interference of electron waves scattered on the defects, which
give rise to small but measurable conductance fluctuations.
Whereas in that work the band structure was assumed to be
free-electron like, here we investigate the effects of Fermi
surface anisotropy. We demonstrate that the amplitude and period
of the conductance oscillations are determined by the local
geometry of the Fermi surface. The signal results from those
points for which the electron velocity is directed along the
vector connecting the point contact to the defect. For a general
Fermi surface geometry the position of the maximum amplitude of
the conductance oscillations is not found for the tip directly
above the defect. We have determined optimal conditions for
determination of defect positions in metals with closed and open
Fermi surfaces.
\end{abstract}

\maketitle

\section{Introduction}

\bigskip The interference of electron waves scattered by single defects
results in an oscillatory dependence of the point contact conductance $%
G\left( V\right) $ on the applied voltage $V.$ This effect originates from
quantum interference between the principal wave that is directly transmitted
through the contact and the partial wave that is scattered by the contact
and the defect or several defects. Such conductance oscillations have been
observed in quantum point contacts \cite{Ludoph1,Untiedt,Ludoph,Kempen} and
investigated theoretically in the papers \cite%
{Ludoph,Namir,Avotina2,Avotina3}.

In our previous paper \cite{Avotina1} the oscillatory voltage
dependence of the conductance of a tunnel point contact in the
presence of a single point-like defect has been analyzed
theoretically and it has been shown that this dependence can be
used for the determination of defect positions below a metal
surface by means of a scanning tunnelling microscope (STM). In the
model of a spherical Fermi surface (FS) the amplitude of the
conductance oscillations is maximal when the contact is placed
directly above the defect. The oscillatory part of the conductance
$\Delta G$ for this situation is proportional to
\begin{equation*}
\Delta G\left( V\right) \thicksim \cos \left( 2z_{0}\sqrt{k_{F}^{2}+\frac{%
2meV}{\hbar ^{2}}}\right) ,
\end{equation*}%
where $z_{0}$ the depth of the defect and $k_{F}$ and $m$ are the Fermi wave
vector and effective mass of the electrons \cite{Avotina1}. Materials with
an almost spherical FS are most suitable for this model.

In most metals the dispersion relation for the charge carriers is
a complicated anisotropic function of the momentum. This leads to
anisotropy of the various
kinetic characteristics \cite{LAK}. Particularly, as shown in Ref.~\cite%
{Kosevich}, the current spreading may be strongly anisotropic in the
vicinity of a point-contact. This effect influences the way the
point-contact conductance depends on the position of the defect. For
example, in the case of a Au(111) surface the `necks' in the FS should cause
a defect to be invisible when probed exactly from above.

Qualitatively, the wave function of electrons injected by a point contact
for arbitrary FS $\varepsilon \left( \mathbf{p} \right) =\varepsilon _{F}$
has been analyzed by A. Kosevich \cite{Kosevich}. He noted that at large
distances from the contact the electron wave function for a certain
direction $\mathbf{r}$ is defined by those points on the FS for which the
electron group velocity is parallel to $\mathbf{r.}$ Unless the entire FS is
convex there are several such points. The amplitude of the wave function
depends on the Gaussian curvature $K$ in these points, which can be convex $%
\left( K>0\right)$ or concave $\left( K<0\right)$. The parts of the FS
having different signs of curvature are separated by lines of $K=0$
(inflection lines). In general there is a continuous set of electron wave
vectors for which $K=0$. The electron flux in the directions having zero
Gaussian curvature exceeds the flux in other directions \cite{Kosevich}.

Electron scattering by defects in metals with an arbitrary FS can be
strongly anisotropic \cite{LAK}. Generally, the wave function of the
electrons scattered by the defect consists of several superimposed waves,
which travel with different velocities. In the case of an open
constant-energy surface there are directions along which the electrons can
not move at all. Scattering events along those directions occur only if the
electron is transferred to a different sheet of the FS \cite{LAK}.

In this paper we analyze the effect of anisotropy of the FS to the
possibility of determination of the position of a defect below a metal
surface by use of a STM. We show that the amplitude and the period of the
conductance oscillations are defined by the local geometry of the FS, namely
by those points for which the electron group velocity is directed along the
radius vector from the contact to the defect. General formulas for the wave
function and point contact conductance are obtained in sections II and III.
In Sec.~IV the asymptotic forms of the wave function and the point contact
conductance for large distances of the defect from the contact are found.
The general results are illustrated for two specific models of the FS: an
ellipsoid (Sec.~V) and a corrugated cylinder (Sec.~VI). Using these models,
for which analytical dependencies of the conductance on voltage and defect
position can be found, we describe the manifestation of common features of
FS geometries to the conductance oscillations: anisotropy of a convex part
(`bellies'), changing of the curvature (inflection lines) and presence of
open directions (`necks').

\section{\protect\bigskip The Schr\"{o}dinger equation for quasiparticles}

Let us consider as a model for our system a nontransparent interface located
at $z=0$ separating two metal half-spaces, in which there is an orifice
(contact) of radius $R$ centered at the point $\mathbf{r}=0$. The potential
barrier in the plane $z=0$ is taken to be a delta function,
\begin{equation}
U\left( \mathbf{r}\right) =Uf\left( \mathbf{\rho }\right) \delta \left(
z\right) ,  \label{U(z)}
\end{equation}
where $\mathbf{\rho }=\left( x,y\right) $ is a two dimensional vector in the
plane of the interface, with $\mathbf{r=}\left( \mathbf{\rho ,}z\right)$.
The function $f\left( \mathbf{\ \rho }\right) \rightarrow \infty $ in all
points of the plane $z=0$ except in the contact, where $f\left( \mathbf{\rho
}\right) =1$. At the point $\mathbf{r}=\mathbf{r}_{0}$ near the contact in
the upper half-space, $z>0$, a point-like defect is placed. The electron
interaction with the defect is described by the potential $D\left( \mathbf{r}%
-\mathbf{r}_{0}\right) ,$ which is confined to a small region with a
characteristic radius $r_{D}$ around the point $\mathbf{r}_{0}$.

It is known that one can obtain an effective Schr\"{o}dinger equation for
quasiparticles in a metal from the dispersion relation $\varepsilon \left(
\mathbf{p}\right) ,$ (the band structure) by replacement of the
quasimomentum $\mathbf{p}$ (below for short we write momentum) in the
function $\varepsilon \left( \mathbf{p}\right) $ with the momentum operator $%
\widehat{\mathbf{p}}=\frac{\hbar }{i}\nabla $ \cite{LAK}. Here we do not
specify the specific form of the dependence $\varepsilon \left( \mathbf{p}%
\right)$, except that it satisfies the general condition of point symmetry $%
\varepsilon \left( \mathbf{p}\right) =\varepsilon \left( -\mathbf{p}\right) $%
. For simplicity we assume that FS has only one sheet; there is only one
zone described by the function $\varepsilon \left( \mathbf{p}\right).$ In
the reduced zone scheme a given momentum $\mathbf{p}$ identifies a single
point within the first Brillouin zone. The wave function $\psi \left(
\mathbf{r }\right) $ satisfies the Schr\"{o}dinger equation with an
effective Hamiltonian $\varepsilon \left( \widehat{\mathbf{p}}\right) $,
\begin{equation}
\varepsilon \left( \widehat{\mathbf{p}}\right) \psi \left( \mathbf{r}\right)
+\left[ \varepsilon -U\left( \mathbf{r}\right) -eV\left( z\right) \right]
\psi \left( \mathbf{r}\right) =D\left( \mathbf{r}-\mathbf{r}_{0}\right) \psi
\left( \mathbf{r}\right) ,  \label{Schrod}
\end{equation}
where $U\left( \mathbf{r}\right) $ is defined by Eq. (\ref{U(z)}),
$V\left( z\right) $ is the applied electrical potential, and
$\varepsilon $ is the electron energy.

We consider a large barrier potential $U.$ In this case the amplitude $t$ of
the electron wave function passing through the barrier is
\begin{equation}
t\left( \varepsilon ,\mathbf{p}_{t}\right) \approx \frac{\hbar \left( v_{z}^{%
\mathrm{in}}-v_{z}^{\mathrm{ref}}\right) }{2iU},  \label{t_prob}
\end{equation}
where $v_{z}^{\mathrm{in}}$ and $v_{z}^{\mathrm{ref}}$ are the $z$%
-components of the velocity $\mathbf{v}=\partial \varepsilon \left( \mathbf{p%
}\right) /\partial \mathbf{p} $ of incident electrons (in) and electrons
specularly reflected by the barrier (ref), respectively. Under condition of
specular reflection the energy $\varepsilon$ and the component of the
momentum tangential to the interface, $\mathbf{p} _{t}=\left(
p_{x},p_{y}\right) $ at $z=0$ are conserved. The components of the electron
momentum perpendicular to the interface, $p_{z}^{\mathrm{in}}\left( \mathbf{p%
}_{t},\varepsilon \right) $ and $p_{z}^{\mathrm{ref}}\left( \mathbf{p}%
_{t},\varepsilon \right) $ are related by the equations,
\begin{equation}
\varepsilon \left( \mathbf{p}_{t}^{\mathrm{in}},p_{z}^{\mathrm{in}}\right)
=\varepsilon \left( \mathbf{p}_{t}^{\mathrm{ref}},p_{z}^{\mathrm{ref}%
}\right) =\varepsilon ;\quad \mathbf{p}_{t}^{\mathrm{in}}=\mathbf{p}_{t}^{%
\mathrm{ref}}\equiv \mathbf{p}_{t}.  \label{pz_ref}
\end{equation}
The velocities $v_{z}^{\mathrm{in}}$ and $v_{z}^{\mathrm{ref}}$ have the
opposite sign
\begin{equation}
\mathbf{v}^{\mathrm{in}}\cdot\mathbf{N}<0,\quad \mathbf{v}^{\mathrm{ref}%
}\cdot\mathbf{N}>0,  \label{v_in-v_ref}
\end{equation}
where $\mathbf{N}$ is a unit vector normal to the interface laying
in the half-space of the electron wave under consideration
($\mathbf{N}=(0,0,1)$ for $z>0$ and $\mathbf{N}=(0,0,-1)$ for
$z<0$). We will assume that the crystallographic axes in
half-spaces $z\gtrless 0$ are identical. In this case the momenta
and velocities for electrons incident on the barrier and for those
transmitted through the barrier are equal.

In general Eq.~(\ref{pz_ref}) may have several solutions, i.e.
several specularly reflected states may correspond to an incident
state with momentum $p_{z}^{\mathrm{in}}.$ Such reflection is
called multichannel specular reflection \cite{Ustinov}. Below we
assume that there is only one reflected electron state.

In the limit of a small probability of electron tunnelling through the
barrier, $\left\vert t\right\vert ^{2}\ll 1$ the applied voltage drops
entirely over the barrier and we take the electric potential to be a step
function $V\left( z\right) =V\,\Theta \left( -z\right) .$ The reference
point of zero electron energy is the bottom of the conduction band in the
upper half-space, $z>0.$ The conduction band in the lower half-space $z<0$
is shifted by a value $eV.$ We also assume that the applied bias $eV$ is
much smaller than the Fermi energy and in solving the Schr\"{o}dinger equation (%
\ref{Schrod}) we neglect the electric potential $V\left( z\right) .$
Equation (\ref{Schrod}) can be solved by using perturbation theory with the
small parameter $\left\vert t\right\vert \ll 1$ \cite{KMO}. In the zeroth
approximation in this parameter we have the problem of an impenetrable
partition between two metal half-spaces.

We start by solving for the wave function $\psi ^{\left( 0\right)
}\left( \varepsilon , \mathbf{p}_{t};\mathbf{r}\right) $ for a
tunnelling point contact of low transparency, $\left\vert
t\right\vert \ll 1$ without
defects ($D\left( \mathbf{r}-\mathbf{r}_{0}\right)=0$). The wave function $%
\psi _{0}^{\left( 0\right) }\left( \mathbf{r} \right) $, in zeroth order in
the parameter $\left\vert t\right\vert \ll 1$, satisfies the boundary
condition $\psi _{0}^{\left( 0\right) }\left( z=0\right) =0$ at the
interface,
\begin{equation}
\psi _{0}^{\left( 0\right) }\left( \varepsilon ,\mathbf{p}_{t};\mathbf{r}
\right) =e^{i\mathbf{p}_{t}\mathbf{\rho}/\hbar}\left( e^{i p_{z}^{\mathrm{in}%
}z/ \hbar}-e^{i p_{z}^{\mathrm{ref}}z/ \hbar}\right) .  \label{psi_0}
\end{equation}%
\qquad

Let us consider an electron wave $\exp \left( i \mathbf{pr}/ \hbar \right) $
incident on the junction from the lower half-space $z<0$, so that $v_{z}>0$.
In this half-space to first approximation in the parameter $t$ the solution $%
\psi ^{\left( 0\right) }\left( \mathbf{r}\right) $ of the homogeneous Schr%
\"{o}dinger equation can be written in the form \cite{KMO}:
\begin{equation}
\psi ^{\left( 0\right) }\left( \mathbf{r}\right) =\psi _{0}^{\left( 0\right)
}\left( \mathbf{r}\right) +\varphi ^{\left( -\right) }\left( \mathbf{r}%
\right) ;\quad z<0,  \label{Psi<0}
\end{equation}%
where the second term, $\varphi ^{\left( -\right) }\left( \mathbf{r}\right)
\propto t$, describes the changes in the reflected wave as a result of
transmission through the contact. The wave function transmitted into the
half-space $z>0$ is proportional to the amplitude $t$,
\begin{equation}
\psi ^{\left( 0\right) }\left( \mathbf{r}\right) =\varphi ^{\left( +\right)
}\left( \mathbf{r}\right) ;\quad z>0.  \label{psi>0}
\end{equation}%
The function $\psi ^{\left( 0\right) }\left( \mathbf{\rho },z\right) $
satisfies the condition of continuity and the condition of conservation of
probability flow at $z=0.$ For small $\left\vert t\right\vert $ these
conditions reduce to
\begin{gather}
\varphi ^{\left( -\right) }\left( \mathbf{\rho },0\right) =\varphi ^{\left(
+\right) }\left( \mathbf{\rho },0\right) ;  \label{bound1} \\
t e^{i \mathbf{p}_{t}\mathbf{\rho }/ \hbar}=f\left( \mathbf{\rho } \right)
\varphi ^{\left( +\right) }\left( \mathbf{\rho },0\right) .
\label{gran_cond}
\end{gather}
In the absence of defects $\left( D=0\right) $ the solution of the Schr\"{o}%
dinger equation is given by \cite{KMO,Avotina1},
\begin{equation}
\varphi ^{\left( +\right) }\left( \mathbf{\rho },z\right)
=t\int\limits_{-\infty }^{\infty }\frac{d\mathbf{p}_{t}^{\prime }}{\left(
2\pi \hbar \right) ^{2}}F\left( \mathbf{p}_{t}-\mathbf{p}_{t}^{\prime
}\right) e^{i\left(\mathbf{p}_{t}^{\prime }\mathbf{\rho }+ p_{z}^{\left(
+\right) }\left( \varepsilon,\mathbf{p}_{t}^{\prime }\right) z\right)/ \hbar
},  \label{psi+00}
\end{equation}
where
\begin{equation}
F\left( \mathbf{p}_{t}-\mathbf{p}_{t}^{\prime }\right) =\int\limits_{-\infty
}^{\infty }d\mathbf{\rho }\frac{e^{i\left( \mathbf{p}_{t}-\mathbf{p}%
_{t}^{\prime }\right) \mathbf{\rho}/\hbar}}{f\left( \mathbf{\rho } \right) };
\label{F}
\end{equation}%
$p_{z}^{\left( \pm \right) }\left( \varepsilon ,\mathbf{p}_{t}^{\prime
}\right) $ are roots of the equation
\begin{equation}
\varepsilon \left( \mathbf{p}_{t}^{\prime },p_{z}^{\left( \pm \right)
}\right) =\varepsilon \left( \mathbf{p}\right) ,  \label{pz+/-}
\end{equation}
corresponding to waves with velocities $v_{z}^{\left( +\right) }\left(
\varepsilon ,\mathbf{p}_{t}^{\prime }\right) >0$ and $v_{z}^{\left( -\right)
}\left( \varepsilon ,\mathbf{p}_{t}^{\prime }\right) <0.$

Let $D\left( \mathbf{r}\right) $ be a spherically symmetric
scattering potential for a point-like defect, with a range $r_{D}$
that is order of the Fermi wave length $\lambda _{\mathrm{F}}$
($D\ll D\left(
0\right) $, the maximal value of $D$, when $\left\vert \mathbf{r}-\mathbf{r}%
_{0}\right\vert \gg r_{D}$). For a point-like defect ($r_{D}\rightarrow 0$),
the right hand side in the Schr\"{o}dinger equation (\ref{Schrod}) can be
rewritten as $D\left( \mathbf{r}-\mathbf{r}_{0}\right) \psi \left( \mathbf{r}%
_{0}\right) $ \cite{Azbel}. This makes it possible to find a solution to Eq.(%
\ref{Schrod}) by means of the Green function $G_{0}^{+}\left( \mathbf{r}%
^{\prime }\mathbf{,r};\varepsilon \right) $ of the homogeneous equation (at $%
D=0$). \ The wave function scattered from the defect, $\psi \left( \mathbf{r}%
\right) $, can be expressed in terms of the wave function $\varphi ^{\left(
+\right) }\left( \mathbf{r}\right) $ transmitted into the upper metal
half-space,
\begin{equation}
\psi \left( \mathbf{r}\right) =\varphi ^{\left( +\right) }\left( \mathbf{r}%
\right) +\varphi ^{\left( +\right) }\left( \mathbf{r}_{0}\right) \frac{%
J\left( \mathbf{r,r}_{0}\right) }{1-J\left( \mathbf{r}_{0},\mathbf{r}%
_{0}\right) },  \label{psi_scat}
\end{equation}%
where
\begin{equation}
J\left( \mathbf{r,r}_{0}\right) =\int d\mathbf{r}^{\prime }D\left( \mathbf{r}%
^{\prime }-\mathbf{r}_{0}\right) G_{0}^{+}\left( \mathbf{r}^{\prime },%
\mathbf{r};\varepsilon \right) .  \label{J_p}
\end{equation}%
Because the Green function has a singularity at $\mathbf{r\rightarrow r}%
^{\prime }$ Eq.~(\ref{psi_scat}) is correct if the integral $J\left( \mathbf{%
r,r}_{0}\right) $ (\ref{J_p}) converges in the point $\mathbf{r}=\mathbf{r}%
_{0}$.

To proceed with further calculations we assume that the scattering
potential is small and use perturbation theory in the interaction
with the defect. This implies that we take $\left\vert J\left(
\mathbf{r}_{0},\mathbf{r} _{0}\right) \right\vert \ll 1.$ The wave
function solution, that is linear in $\left\vert t\right\vert $
and with a contribution to first order in $D$, is
\begin{equation}
\psi \left( \mathbf{r}\right) =\varphi ^{\left( +\right) }\left( \mathbf{r}
\right) +\varphi ^{\left( +\right) }\left( \mathbf{r}_{0}\right) \int d
\mathbf{r}^{\prime }D\left( \mathbf{r}^{\prime }-\mathbf{r}_{0}\right)
G_{0}^{+}\left( \mathbf{r,r}^{\prime };\varepsilon \right) .
\label{psi_approx}
\end{equation}%
The Green function $G_{0}\left( \mathbf{r,r}^{\prime };\varepsilon
\right) $ in zeroth approximation in the parameters $D$ and
$\left\vert t\right\vert $ should be calculated from the wave
functions (\ref{psi_0}),
\begin{equation}
G_{0}^{+}\left( \mathbf{r,r}^{\prime };\varepsilon \right) =\int \frac{d
\mathbf{p}}{\left( 2\pi \hbar \right) ^{3}}\frac{\psi _{0}^{\left( 0\right)
}\left( \mathbf{r}\right) \left( \psi _{0}^{\left( 0\right) }\left( \mathbf{%
r }^{\prime }\right) \right) ^{\ast }}{\varepsilon -\varepsilon \left(
\mathbf{\ p}\right) -i0}.  \label{G_funct}
\end{equation}
This is the Green function of the Schr{\"{o}}dinger equation in the
half-space $z>0$ with hard-wall boundary conditions at $z=0$. Substituting
the wave function $\psi_{0}^{\left(0\right)}\left( \mathbf{r}\right)$, Eq.~(%
\ref{psi_0}), for $z>0$ in Eq.(\ref{G_funct}) we find,
\begin{equation}
G_{0}^{+}\left( \mathbf{r,r}^{\prime };\varepsilon \right) =\frac{1}{i\hbar}
\int \frac{d\mathbf{p}_{t}}{\left( 2\pi \hbar \right)^{2}} \frac{e^{i
\mathbf{p}_{t}\left( \mathbf{\rho -\rho }^{\prime }\right)/ \hbar + i
p_{z}^{\left( +\right) }z/ \hbar}}{v_{z}^{\left( +\right) }-v_{z}^{\left(
-\right) }} \left( e^{-i p_{z}^{\left( +\right) }z^{\prime} / \hbar} - e^{-i
p_{z}^{\left( -\right) }z^{\prime } / \hbar}\right) ;z>z^{\prime }.
\label{Gz>z0}
\end{equation}%
For $0<z<z^{\prime }$ one should make the replacements $z\leftrightarrow
z^{\prime }$ and $p_{z}^{\left( -\right) }\leftrightarrow -p_{z}^{\left(
+\right) }$ in Eq.~(\ref{Gz>z0}); $p_{z}^{\left( \pm \right) }$ are given by
Eq.~( \ref{pz+/-}).

The main contribution to the integral in Eq.~(\ref{psi_approx}) comes from a
small region near the point $\mathbf{r}^{\prime }=\mathbf{r}_{0}$. Far from
the point $\mathbf{r}=\mathbf{r}_{0}$ ($\left\vert \mathbf{r}-\mathbf{r}%
_{0}\right\vert \gg r_{D})$ the solution (\ref{psi_approx}) takes the form:
\begin{equation}
\psi \left( \mathbf{r}\right) =\varphi ^{\left( +\right) }\left( \mathbf{r}
\right) +gG_{0}^{+}\left( \mathbf{r,r}_{0};\varepsilon \right) \varphi
^{\left( +\right) }\left( \mathbf{r}_{0}\right) ,  \label{psi0+g}
\end{equation}
where
\begin{equation}
g=\int d\mathbf{r}^{\prime }D\left( \mathbf{r}^{\prime }-\mathbf{r}
_{0}\right)
\end{equation}
is the constant of electron-impurity interaction.

\section{Point-contact conductance}

\bigskip The electrical current $I\left( V\right) $ can be evaluated from
the electron wave functions, $\psi $, of the system through \cite{ISh},
\begin{equation}
I\left( V\right) =\frac{2e}{\left( 2\pi \hbar \right) ^{3}}\int d\mathbf{p}%
I_{\mathbf{p}}\Theta \left( v_{z}\right) \left[ n_{\mathrm{F}}\left(
\varepsilon \right) -n_{\mathrm{F}}\left( \varepsilon +eV\right) \right] .
\label{current}
\end{equation}%
Here%
\begin{equation}
I_{\mathbf{p}}=\int\limits_{-\infty }^{\infty }dxdy\ \mathrm{\func{Re}}%
\left( \psi ^{\ast }\widehat{v}_{z}\psi \right)  \label{den_flow}
\end{equation}%
is the density of probability flow in the $z$ direction for the momentum $%
\mathbf{p,}$ integrated over a plane $z=$~xconst,
$n_{\mathrm{F}}\left(
\varepsilon \right) $ is the Fermi distribution function, and $\widehat{%
\mathbf{v}}$ is the velocity operator, $\widehat{\mathbf{v}}=\partial
\varepsilon \left( \widehat{\mathbf{p}}\right) /\partial \widehat{\mathbf{p}}%
.$ For the definiteness we choose in Eq.(\ref{current}) $eV>0.$ At low
temperatures the tunnel current is due to those electrons in the half-space $%
z<0$ having an energy between the Fermi energy, $\varepsilon _{\mathrm{F}},$
and $\varepsilon _{\mathrm{F}}+eV$, because on the other side of the barrier
$z>0$ only states with $\varepsilon \geqslant \varepsilon _{\mathrm{F}}$ are
available.

After performing the integration over a plane at $z\gg z_{0},$ where the
wave function (\ref{psi0+g}) can be used, we find the density flow (\ref%
{den_flow}) becomes,
\begin{gather}
I_{\mathbf{p}}=\left\vert t\left( \varepsilon ,\mathbf{p}_{t}\right)
\right\vert ^{2}\pi ^{3}\hbar R^{4}\left\langle \left( v_{z}^{\left(
+\right) }\right) ^{2}\right\rangle _{\varepsilon }\nu \left( \varepsilon
\right) +  \label{flow_int} \\
\frac{\left\vert t\left( \varepsilon ,\mathbf{p}_{t}\right) \right\vert
^{2}g\pi ^{2}R^{4}}{\hbar }\func{Re}\int \frac{d\mathbf{p}_{t}^{\prime }}{%
\left( 2\pi \hbar \right) ^{2}}\frac{iv_{z}^{\left( +\right) }\left( \mathbf{%
p}_{t}^{\prime }\right) e^{-i\mathbf{p}_{t}^{\prime }\mathbf{\rho }%
_{0}/\hbar }}{v_{z}^{\left( +\right) }\left( \mathbf{p}_{t}^{\prime }\right)
-v_{z}^{\left( -\right) }\left( \mathbf{p}_{t}^{\prime }\right) }\left(
e^{-ip_{z}^{\left( +\right) }\left( \mathbf{p}_{t}^{\prime }\right)
z_{0}/\hbar }-e^{-ip_{z}^{\left( -\right) }\left( \mathbf{p}_{t}^{\prime
}\right) z_{0}/\hbar }\right) \times  \notag \\
\int\limits_{-\infty }^{\infty }\frac{d\mathbf{p}_{t}^{\prime \prime }}{%
\left( 2\pi \hbar \right) ^{2}}e^{i\left( \mathbf{p}_{t}^{\prime \prime }%
\mathbf{\rho }_{0}+p_{z}^{\left( +\right) }\left( \mathbf{p}_{t}^{\prime
\prime }\right) z_{0}\right) /\hbar };  \notag
\end{gather}%
where the mean value at energy $\varepsilon $ is obtained by integration
over the surfaces of constant energy in momentum space,
\begin{equation}
\left\langle \dots \right\rangle _{\varepsilon }=\frac{\int\limits_{%
\varepsilon _{\mathbf{p}}=\varepsilon }\frac{dS_{\mathbf{p}}}{\left\vert
\mathbf{v}\right\vert }\dots }{\int\limits_{\varepsilon _{\mathbf{p}%
}=\varepsilon }\frac{dS_{\mathbf{p}}}{\left\vert \mathbf{v}\right\vert }},
\label{int_en}
\end{equation}%
scaled by the velocity, $\left\vert \mathbf{v}\right\vert =\left\vert \frac{%
\partial \varepsilon }{\partial \mathbf{p}}\right\vert $. In Eq.~(\ref%
{flow_int}), $p_{z}^{\left( +\right) }\left( \mathbf{p}_{t}\right) $ and $%
p_{z}^{\left( -\right) }\left( \mathbf{p}_{t}\right) $ are given by Eq.~(\ref%
{pz+/-}) and $\nu \left( \varepsilon \right) $ is the electron density of
states per unit volume.

Taking into account Eq.~(\ref{flow_int}) we can calculate the
current-voltage characteristics $I\left( V\right) $. The conductance $%
G\left( V\right) $ is the first derivative of the current $I\left( V\right) $
with $V$ and in the limit $T\rightarrow 0$ we obtain,
\begin{equation}
G\left( V\right) =\frac{\partial I}{\partial V}=e^{2}\nu \left( \varepsilon
\right) \left\langle I_{\mathbf{p}}\right\rangle _{\varepsilon _{\mathrm{F}%
}+eV}.  \label{conduct}
\end{equation}%
After integration over $S_{\mathbf{p}}$ (\ref{int_en}),
Eq.(\ref{conduct}) should be expanded in the parameter
$eV/\varepsilon _{\mathrm{F}}\ll 1.$

\section{Asymptotics of the wave function and the conductance}

In this section we find the wave function at a large distances from the
contact, $r\gg \lambda _{\mathrm{F}}$, and an asymptotic expression for the
conductance in the limit of a large distance between the defect and the
contact, $r_{0}\gg \lambda _{\mathrm{F}}$ and a small contact radius, $R\ll
\lambda _{\mathrm{F}}$, where $\lambda _{\mathrm{F}}$ is the characteristic
electron Fermi wave length. For $R\rightarrow 0$ the function $F$ in Eq.~(%
\ref{F}) takes the form \cite{Avotina1},
\begin{equation}
F\left( \mathbf{p}_{t}-\mathbf{p}_{t}^{\prime }\right) =\pi R^{2},
\label{Fa->0}
\end{equation}%
and the wave function (\ref{psi+00}) can be written as
\begin{equation}
\varphi ^{\left( +\right) }\left( \mathbf{\rho },z\right)
=t\left(\varepsilon ,\mathbf{p}_{t}\right) \pi R^{2}\int\limits_{-\infty
}^{\infty }\frac{d\mathbf{p}_{t}^{\prime}}{\left( 2\pi \hbar \right) ^{2}}%
e^{i\left(\mathbf{p}_{t}^{\prime}\mathbf{\rho} + p_{z}^{\left(
+\right)}\left(\mathbf{p}_{t}^{\prime }\right) z\right)/\hbar}.
\label{phi+0}
\end{equation}

Let us consider the integral
\begin{equation}
\Lambda \left( \mathbf{r,}\varepsilon \right) =
\int\limits_{-\infty}^{\infty }\frac{d\mathbf{p}_{t}^{\prime }}{\left( 2\pi
\hbar \right) ^{2}}e^{i\Gamma \left( \mathbf{p}_{t},\mathbf{r}\right) },
\label{lambda}
\end{equation}%
where $\Gamma \left( \mathbf{p}_{t},\mathbf{r}\right) $ is the phase
accumulated over the path travelled by the electron between the contact and
the point $\mathbf{r}$,
\begin{equation}
\Gamma \left( \mathbf{p}_{t},\mathbf{r}\right) =\frac{1}{\hbar }\left(
\mathbf{p}_{t}\mathbf{\rho }+p_{z}^{\left( +\right) }\left( \mathbf{p}%
_{t}\right) z\right) .  \label{phase}
\end{equation}%
This kind of integrals appear in the expressions for the wave function
(Eqs.~(\ref{phi+0}), and (\ref{Gz>z0}),(\ref{psi0+g})) and for the
conductance (Eqs.~(\ref{flow_int}),(\ref{conduct})). At a large distance, $%
r\gg \lambda_{\mathrm{F}}$, the exponent under the integral in Eq.~(\ref%
{lambda}) is a rapidly oscillating function and the integral can be
calculated by the stationary phase method (see, for example, Ref.~\cite%
{Fedoriuk}). The stationary phase points $\mathbf{p}_{t}=\mathbf{p}
_{t}^{\left( \mathrm{st}\right ) }$ are defined by the equation,
\begin{equation}
\left. \frac{\partial \Gamma }{\partial \mathbf{p}_{t}}\right\vert _{\mathbf{%
p}_{t}=\mathbf{p}_{t}^{\left( \mathrm{st}\right) }}=0.  \label{st_point}
\end{equation}%
With Eq.~(\ref{phase}) we find:%
\begin{equation}
\mathbf{\rho } + z\left. \frac{\partial p_{z}^{\left( +\right) }\left(
\mathbf{p}_{t}\right) }{\partial \mathbf{p}_{t}}\right\vert _{\mathbf{p}_{t}=%
\mathbf{p}_{t}^{\left( \mathrm{st}\right) }}=\mathbf{\rho } - z\left. \frac{%
\mathbf{v}_{t}^{\left( +\right) }\left( \mathbf{p}_{t}\right) }{v_{z}\left(
\mathbf{p}_{t}\right) }\right\vert _{\mathbf{p}_{t}=\mathbf{p}_{t}^{\left(
\mathrm{st}\right) }}=0.  \label{st_point2}
\end{equation}
For $r\gg \lambda _{\mathrm{F}}$ the asymptotic value $\Lambda ^{\mathrm{as}%
}\left( \mathbf{r,} \varepsilon \right) $ of the integral (\ref{lambda}) is
given by
\begin{equation}
\Lambda ^{\mathrm{as}}\left( \mathbf{r,}\varepsilon \right) =\frac{\cos
\vartheta }{2\pi \hbar r\sqrt{\left\vert K_{0}\right\vert }}\exp \left[
i\Gamma _{0}+i\frac{\pi }{4}\text{sgn}\left( \frac{\partial ^{2}p_{z}}{%
\partial p_{x}^{\left( \mathrm{st}\right) 2}}\right) \left( 1+\text{sgn}%
K_{0}\right) \right] .  \label{lam}
\end{equation}%
Here%
\begin{equation}
\Gamma _{0}\left( \varepsilon ,\mathbf{r}\right) =\Gamma \left( \mathbf{p}%
_{t}^{\left( \mathrm{st}\right) },\mathbf{r}\right)
\end{equation}%
is the phase (\ref{phase}) in a point defined by Eq.~(\ref{st_point2}), $%
K\left( \varepsilon ,\mathbf{p}\right) $ is the Gaussian curvature
of the surface of constant energy $\varepsilon \left(
p_{x},p_{y},p_{z}\right) =\varepsilon$, and $\cos\vartheta(
\mathbf{r}) =z/r$ is the angle between the vector $\mathbf{r}$ and
the $z$ axis. At the stationary phase points the curvature
$K\left(\varepsilon, \mathbf{p} \right) $ can be written as
\begin{equation}
K_{0}\left( \varepsilon ,\mathbf{n}\right) =\left[
\frac{1}{\left\vert \mathbf{v}\right\vert
^{2}}\sum_{i,k=x,y,z}A_{ik}n_{i}n_{k}\right] _{\mathbf{
p}_{t}=\mathbf{p}_{t}^{\left( \mathrm{st}\right) }},
\label{Gaus0}
\end{equation}
where $A_{ik}=\frac{\partial \det \left( \mathbf{m}^{-1}\right)
}{\partial m_{ik}^{-1}\left( \mathbf{p}\right) }$ is the algebraic
adjunct of the element
\begin{equation}
m_{ik}^{-1}\left( \mathbf{p}\right) =\frac{\partial ^{2}\varepsilon }{%
\partial p_{i}\partial p_{k}}  \label{m^(-1)}
\end{equation}%
of the inverse mass matrix $\mathbf{m}^{-1}$ \cite{Korn}; $n_{i}$
are components of the unit vector $\mathbf{n}=\mathbf{r}/r.$ Note
that for an
arbitrary FS $m_{ik}( \mathbf{p})$ in the point $\mathbf{p}_{t}=\mathbf{p}%
_{t}^{\left( \mathrm{st}\right) }$ depends on the direction of vector $%
\mathbf{r}$. It is follows from Eq.~(\ref{st_point2}) that the velocity at
the stationary phase point $\mathbf{p}_{t}^{\left( \mathrm{st}\right) }$is
parallel to the radius-vector $\mathbf{r=}\left( \mathbf{\rho ,}z\right)$.

If the curvature of the FS changes sign, Eq.~(\ref{st_point2}) has more then
one solution $\mathbf{p}_{t}=\mathbf{p}_{t,s}^{\left( \mathrm{st}\right) }$ (%
$s=1,2...$). In that case the value of the integral (\ref{lambda}) is
replaced by a sum over all points $\mathbf{p}_{t,s}^{\left( \mathrm{st}%
\right) }$, which in the limit of large distances is,
\begin{equation}
\Lambda \left( \mathbf{r,}\varepsilon \right) \approx \sum\limits_{s}\Lambda
_{s}^{\mathrm{as}}\left( \mathbf{r,}\varepsilon \right) ,
\end{equation}%
with $\Lambda _{s}^{\mathrm{as}}\left( \mathbf{r,}\varepsilon
\right) $ given by Eq.~(\ref{lam}) for each stationary phase point
$s$. It may also occur that Eq.~(\ref{st_point2}) does not have
any solution for given directions of the vector $\mathbf{r}$, and
the electron cannot propagate along these directions. These two
energy surface properties result in complicated patterns of the
distribution of the modulus of the wave function: 1) For
directions for which Eq.~(\ref{st_point2}) has several solutions a
quantum interference pattern of the electron waves with different
velocities should be observed. 2) When Eq.~(\ref{st_point2}) has
no solution for the selected direction of the vector $\mathbf{r}$
classical motion in this direction is forbidden and the wave
function is exponentially small.

For large values $r,r^{\prime }\gg \lambda _{\mathrm{F}}$ the asymptotic
behaviors of the Green function, Eq.~(\ref{Gz>z0}), and of the conductance,
Eq.~(\ref{conduct}) at $r_{0}\gg \lambda _{\mathrm{F}}$, can be found
analogous to the evaluation of the integral in Eq.~(\ref{lambda}). The
slowly varying functions of the momentum must be taken in the stationary
phase point. For the partial wave scattered by the defect that is moving
towards the interface $p_{z}^{\left( +\right) }\left( \mathbf{p}_{t}\right) $
in Eq.~(\ref{phase}) must be replaced by $p_{z}^{\left( -\right) }\left(
\mathbf{p}_{t}\right)$. In this case the stationary points have a group
velocity $\mathbf{v}^{\left( -\right) }$ directed from the point $\mathbf{r}%
_{0}$ towards the contact. From the central symmetry of the FS, $\varepsilon
\left( \mathbf{p}\right) =\varepsilon \left( -\mathbf{p}\right) $, it
follows that the two stationary phase points for the function $\Gamma \left(
\mathbf{p}_{t},\mathbf{r}\right) $ are antiparallel, $\mathbf{p}^{\left(
+\right) \left( \mathrm{st}\right) }=-\mathbf{p}^{\left( -\right) \left(
\mathrm{st}\right) }\equiv \mathbf{p}^{\left( \mathrm{st}\right) }$.

Next, we derive an asymptotic expression for the wave function (\ref{psi0+g}%
), with $r, \left\vert \mathbf{r-r}_{0}\right\vert \gg \lambda _{\mathrm{F}}$%
, for a symmetric orientation of the FS with respect to the interface, so
that $v_{z}^{\left( +\right) }=-v_{z}^{\left( -\right) }$, $p_{z}^{\left(
+\right) }=-p_{z}^{\left( -\right) }$ and $m_{ik}^{-1}=0$, if $i\neq k$.
Under these conditions the wave function (19) takes the form,
\begin{eqnarray}
\psi \left( \mathbf{r}\right) &\thickapprox &\varphi^{\left( +\right)}\left(%
\mathbf{r}\right) + \frac{gi}{4\pi\hbar}\varphi^{\left( +\right)}\left(%
\mathbf{r}_{0}\right) \sum\limits_{s}\left( \frac{1}{v_{z}^{\left( +\right)}(%
\mathbf{p}_{t,s}^{\left( 1\right)}) } \Lambda _{s}^{\mathrm{as}}\left(%
\mathbf{\rho-\rho}_{0},z+z_{0};\varepsilon \right) \right. -  \notag \\
&&\left. \frac{1}{v_{z}^{\left( +\right) }(\mathbf{p}_{t,s}^{\left( 2\right)
}) } \Lambda _{s}^{\mathrm{as}}\left( \mathbf{\rho -\rho }_{0},\left\vert
z-z_{0}\right\vert ;\varepsilon \right) \right),  \label{psi_asym}
\end{eqnarray}%
with%
\begin{equation}
\varphi ^{\left( +\right) }\left( \mathbf{r}\right) \thickapprox t\pi
R^{2}\sum_{s}\Lambda _{s}^{\mathrm{as}}\left( \mathbf{r,}\varepsilon \right)
.  \label{psi0_asym}
\end{equation}%
In Eq.~(\ref{psi_asym}) the velocity $v_{z}^{\left( +\right) }$ is taken in
the stationary phase points $\mathbf{p}_{t}^{(st)}=\mathbf{p}_{t,s}^{\left(
1,2\right) }$ corresponding to the directions of the vector with coordinates
$\left( \mathbf{\rho -\rho }_{0},\left\vert z\pm z_{0}\right\vert \right)$.

We have assumed for the Gaussian curvature, Eq.~(\ref{Gaus0}), that $%
K_{0}\neq 0$ in the stationary phase points $\mathbf{p}_{t}^{\left( \mathrm{%
st}\right) }$. For those points at which $K_{0}=0$ the integral (\ref{lambda}%
) diverges. This means that the third derivative of the phase $\Gamma \left(
\mathbf{p}_{t},\mathbf{r}\right) $ (\ref{phase}) with respect to $\mathbf{p}%
_{t}$ must be taken into account. In Sec.~V this is done for a model FS
having cylindrical symmetry with respect to an open direction. Here we only
note that the amplitude of the wave function $\varphi ^{\left( +\right)
}\left( \mathbf{r}\right) $ (\ref{psi0_asym}) in a direction of zero
Gaussian curvature is larger than for other directions, and decreases more
slowly as compared to the $\thicksim 1/r$ dependence of Eq.(\ref{lam}). This
results in an enhanced current flow near the cone surface defined by the
condition $K_{0}=0$ \cite{Kosevich}. The same effect also appears in the
second term of the wave function (\ref{psi_asym}), that describes the
scattered wave: the amplitude of this partial wave is maximal in the
directions of zero Gaussian curvature.

If the FS has a flat part, i.e. Eq.(\ref{st_point2}) holds at all
points of a FS region of finite area $S_{\mathrm{fl}}$, the
associated electron waves propagate in the metal without decrease
of their amplitude \cite{Kosevich}. For the flat part the
dispersion relation $\varepsilon
\left( \mathbf{p}\right) $ can be presented as $\varepsilon \left( \mathbf{p}%
\right) =\mathbf{v}_{0}\mathbf{p,}$ where $\mathbf{v}_{0}={\rm
const}$ is the
electron velocity. For such FS the asymptotic value of the integral (\ref%
{lambda}) is%
\begin{equation}
\Lambda _{\mathrm{fl}}^{\mathrm{as}}\left( \mathbf{r,}\varepsilon \right) =%
\frac{v_{0z}S_{\mathrm{fl}}}{\left\vert \mathbf{v}_{0}\right\vert \left(
2\pi \hbar \right) ^{2}}\exp \left( \frac{i\varepsilon r}{\hbar \left\vert
\mathbf{v}_{0}\right\vert }\right) .  \label{lambda_fl}
\end{equation}

When the distance between the contact and the defect is large,
$r_{0}\gg \lambda _{\mathrm{F}}$, we obtain the conductance of the
tunnel junction, using Eq.~(\ref{flow_int}) and the asymptotic
expression for the wave function (\ref{psi_asym}),
\begin{equation}
G=G_{0}\left( 1-\frac{g}{2\pi \hbar ^{2}\left\langle \left( v_{z}^{\left(
+\right) }\right) ^{2}\right\rangle _{\varepsilon _{\mathrm{F}}}\nu \left(
\varepsilon _{\mathrm{F}}\right) }\sum_{s,s^{\prime }}\func{Re}\Lambda _{s}^{%
\mathrm{as}}\left( \mathbf{r}_{0}\mathbf{,}\varepsilon _{\mathrm{F}%
},eV\right) \func{Im}\Lambda _{s^{\prime }}^{\mathrm{as}}\left( \mathbf{r}%
_{0}\mathbf{,}\varepsilon _{\mathrm{F}},eV\right) \right) ,
\label{G_asym}
\end{equation}%
where $G_{0}$ is the zero-bias ($eV\rightarrow 0$) conductance of
the junction without defect ,
\begin{equation}
G_{0}=e^{2}\pi ^{3}R^{4}\hbar \left\langle \left\vert t\right\vert
^{2}\right\rangle _{\varepsilon _{\mathrm{F}}}\nu ^{2}(\varepsilon _{\mathrm{%
F}})\left\langle \left( v_{z}^{\left( +\right) }\right) ^{2}\right\rangle
_{\varepsilon _{\mathrm{F}}}.  \label{G0_arbit}
\end{equation}
In deriving Eq.~(\ref{del_G}) we have assumed that $eV\ll \varepsilon _{%
\mathrm{F}}$ and $r_{0}\gg \lambda _{\mathrm{F}}.$ Therefore, all
functions
of the energy $\varepsilon $ in Eq.~(\ref{del_G}) can be taken at $%
\varepsilon =\varepsilon _{\mathrm{F}}$, except for the phase
$\Gamma
_{0}\left( \varepsilon ,\mathbf{r}\right) $. When $eV\ll \varepsilon _{%
\mathrm{F}}$,
\begin{equation}
\Gamma _{0}\left( \mathbf{r}_{0},\varepsilon
_{\mathrm{F}}+eV\right) \approx \Gamma _{0}\left( \varepsilon
_{\mathrm{F}}\right) +\frac{\partial \Gamma
_{0}}{\partial \varepsilon _{\mathrm{F}}}eV,\quad \frac{\partial \Gamma _{0}%
}{\partial \varepsilon _{\mathrm{F}}}\sim \frac{1}{\varepsilon _{\mathrm{F}}}%
\frac{r_{0}}{\lambda _{\mathrm{F}}}
\end{equation}%
and when the product $(eV/\varepsilon _{\mathrm{F}})(r_{0}/\lambda _{\mathrm{%
F}})\gg 1$ clearly the conductance (\ref{G_asym}) is an
oscillatory function
of the voltage $V$. Note that if the inequality $eV\ll \varepsilon _{\mathrm{%
F}}$ is not satisfied, the value of the conductance
(\ref{G0_arbit}) as well as the amplitude of its oscillations
depend on $V.$ The periods of oscillations are defined by the
energy dependence of the function $\Gamma _{0}\left( \varepsilon
,\mathbf{r}_{0}\right) $ and they remain the same for any voltage.
Below presenting formulas for the conductance in different
cases we don't expand arguments of oscillatory functions in the parameter $%
eV/\varepsilon _{\mathrm{F}}.$ The obtained results properly
describe the total conductance at $eV\ll \varepsilon
_{\mathrm{F}}$ and also can be used
for the analysis of periods of oscillations at $eV\leq \varepsilon _{\mathrm{%
F}}.$ In Eq.(\ref{G_asym}) the $\Lambda _{s}^{\mathrm{as}}\left( \mathbf{r}%
_{0}\mathbf{,}\varepsilon _{\mathrm{F}},eV\right) $ denotes the function (%
\ref{lam}), in which  $\Gamma _{0}=\Gamma _{0}\left( \mathbf{r}%
_{0},\varepsilon _{\mathrm{F}}+eV\right) .$ Equation
(\ref{G_asym}) shows that for a tunnel junction of small size the
amplitude and period of conductance oscillations depend on the
local geometry FS in those points for which the velocity is
directed along the vector $\mathbf{r}_{0}$.

In the case of a convex FS there is only one stationary phase point
satisfying Eq.~(\ref{st_point2}) which allows simplifying the expression for
the conductance (\ref{G_asym}). The oscillating part of the conductance, $%
\Delta G$, can be written as,
\begin{equation}
\frac{\Delta G}{G_{0}}=\frac{g}{2\left\langle \left( v_{z}^{\left( +\right)
}\right) ^{2}\right\rangle _{\varepsilon _{\mathrm{F}}}\nu \left(
\varepsilon _{\mathrm{F}}\right) \left( 2\pi \hbar \right) ^{3}r_{0}^{2}}%
\left[ \frac{\left( v_{z}^{\left( +\right) }\left( \varepsilon ,\mathbf{p}%
_{t}\right) \right) ^{2}}{\sum\limits_{i,k=x,y,z}A_{ik}n_{0i}n_{0k}}\sin
\left( 2\Gamma _{0}(\mathbf{r}_{0},\varepsilon _{\mathrm{F}}+eV)\right) %
\right] _{\mathbf{p}_{t}=\mathbf{p}_{t}^{\left( \mathrm{st}\right)
}}, \label{del_G}
\end{equation}%
where $\mathbf{n}_{0}=\mathbf{r}_{0}/r_{0}$. The phase of the
oscillations in the conductance, $2\Gamma _{0}$, is determined by
the phase that the electron accumulates along the trajectory from
the contact to the defect and back
\begin{equation}
2\Gamma _{0}\left( \mathbf{r}_{0},\varepsilon \right) =\frac{2}{\hbar }%
\mathbf{p}^{\left( \mathrm{st}\right) }\mathbf{r}_{0},
\label{Ga_el}
\end{equation}%
where $2\left\vert \mathbf{p}^{\left( \mathrm{st}\right) }\right\vert $ is
the chord connecting the two points on the surface of constant energy for
which the velocities are antiparallel and aligned with the vector $\mathbf{r}%
_{0}$.

If the direction from the contact to the defect coincides with a
direction of electron velocities for a flat part of the FS we
should use asymptotic expression (\ref{lambda_fl}) for the
function $\Lambda _{\mathrm{fl}}^{\mathrm{as}}(
\mathbf{r}_{0}\mathbf{,}\varepsilon )$ when calculating the
conductance (\ref{G_asym}).  For this case the oscillating part of
the conductance, $\Delta G$, is given by,
\begin{equation}
\frac{\Delta G}{G_{0}}=\frac{\pi gS_{\mathrm{fl}}^{2}v_{z0}^{2}}{%
\left\langle \left( v_{z}^{\left( +\right) }\right) ^{2}\right\rangle
_{\varepsilon _{\mathrm{F}}}\left\vert \mathbf{v}_{0}\right\vert ^{2}\nu
\left( \varepsilon _{\mathrm{F}}\right) \left( 2\pi \hbar \right) ^{4}}\sin
\left( \frac{2\left( \varepsilon _{\mathrm{F}}+eV\right) r_{0}}{\hbar
\left\vert \mathbf{v}_{0}\right\vert }\right) .
\end{equation}%
Note that $r_{0}/\left\vert \mathbf{v}_{0}\right\vert
=z_{0}/v_{z0}$. In the case when the FS has a flat part, the
amplitude of the conductance oscillations for those special
directions does not depend on the distance between the contact and
the defect.

In the next sections we shall consider two models of anisotropic
FSs. The first model, of an ellipsoidal FS, illustrates the main
features of the conductance oscillations for metals with a convex
FS, i.e. with positive Gaussian curvature. The FS of the second
model has the shape of a corrugated cylinder, which illustrates
the effects of sign inversion of the curvature and the presence of
open directions of the FS. These two models allow us to obtain
dependencies of the conductance on the applied voltage and on the
defect position in analytical form.

\section{Ellipsoidal Fermi surface}

For an ellipsoidal FS the Schr\"{o}dinger equation (\ref{Schrod}) can, in
fact, be solved exactly in the limit $R\rightarrow 0$ and the wave function (%
\ref{psi0+g}) and the conductance (\ref{conduct}) can be found for
arbitrary distances between the contact and the defect. For this
FS the dependence of the electron energy $\varepsilon $ on the
momentum $\mathbf{p}$ is given by relation,
\begin{equation}
\varepsilon \left( \mathbf{p}\right) =\frac{1}{2}\sum_{i,k=x,y,z}\frac{%
p_{k}p_{i}}{m_{ik}};  \label{energy}
\end{equation}%
were $p_{i}$ are the components of the electron momentum $\mathbf{p}$, $%
1/m_{ik}$ are constants representing the components of the inverse effective
mass tensor $\mathbf{m}^{-1}$. The tensor $\mathbf{m}^{-1}$ can be
diagonalized to the form $\left\{ \mathbf{m}^{-1}\right\}
_{ik}=m_{i}^{-1}\delta _{ik}$ so that the momentum-space lengths $\sqrt{%
2\varepsilon _{\mathrm{F}}m_{i}}$ correspond the semi-axes of the FS
ellipsoid $\varepsilon \left( \mathbf{p}\right) =\varepsilon _{\mathrm{F}}$.

\begin{figure}[tbp]
\includegraphics[width=8cm,angle=0]{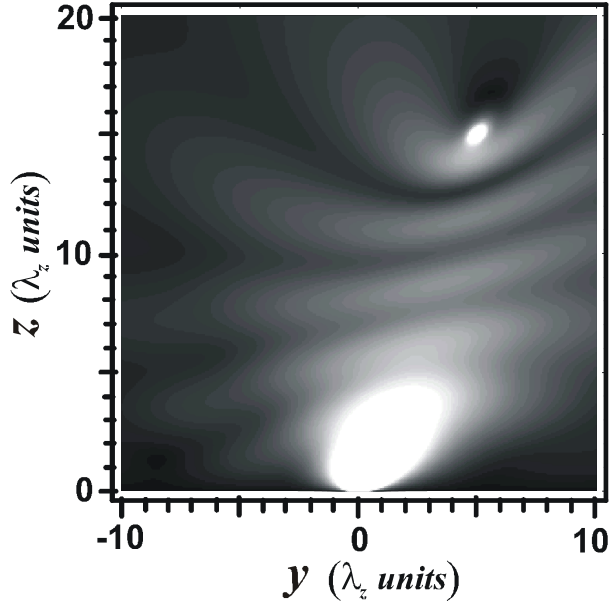}
\caption{Gray-scale plot of the modulus of the wave function in
the plane $x=0$ for an ellipsoidal
FS. The shape of the FS is defined by the mass ratios $m_{x}/m_{z}$=1, $%
m_{y}/m_{z}=3$, and the long axis of the ellipsoid is rotated by $\protect%
\pi /4$ around the $x-$axis, away from the $y-$axis. The
coordinates are measured in units of $\lambda _{z}=\hbar
/\protect\sqrt{2m_{zz}\protect\varepsilon }$. The position of the
defect is $\mathbf{r}_{0}=(0,5,15)$. } \label{fig-ell-abs-phi}
\end{figure}

For the ellipsoidal FS in absence of a defect (zeroth approximation) the
wave function $\varphi ^{\left( +\right) }\left( \mathbf{r}\right) $ (\ref%
{phi+0}) can be obtained by integration of Eq.~(\ref{phi+0}) over momentum,
\begin{equation}
\varphi ^{\left( +\right) }\left( \mathbf{r}\right) =t\left( \varepsilon ,%
\mathbf{p}_{t}\right) \frac{z\pi R^{2}\left( 2\varepsilon \right) ^{3/2}}{%
\hbar ^{3}\sqrt{\det \mathbf{m}^{-1}}}\frac{e^{i\Gamma _{0}}\left( 1-i\Gamma
_{0}\left( \mathbf{r}\right) \right) }{\Gamma _{0}^{3}\left( \mathbf{r}%
\right) },  \label{phi0_ell}
\end{equation}%
where%
\begin{equation}
\Gamma _{0}\left( \varepsilon ,\mathbf{r}\right) =\frac{r}{\hbar }\sqrt{%
2\varepsilon \sum_{i,k=x,y,z}\mu _{ik}n_{i}n_{k}},
\end{equation}%
and
\begin{equation}
t\left( \varepsilon ,\mathbf{p}_{t}\right) =\frac{\hbar }{iUm_{zz}}\sqrt{%
2m_{zz}\varepsilon -m_{zz}\sum\limits_{i,k=x,y}\frac{p_{i}p_{k}}{m_{ik}}%
+\left( m_{zz}\sum\limits_{i=x,y}\frac{p_{i}}{m_{zi}}\right) ^{2}}.
\end{equation}%
Here, $\mu _{ik}$ are the elements of the mass tensor $\mathbf{\mu }$, which
is the inversive to the tensor $\mathbf{m}^{-1}$ ($\mathbf{\mu m}^{-1}=%
\mathbf{I}$, with $\mathbf{I}$ the unitary tensor; $\mu _{ik}=A_{ik}/\det
\mathbf{m}^{-1}$). The Green function (\ref{Gz>z0}) takes the form
\begin{equation}
G_{0}^{+}\left( \mathbf{r,r}^{\prime };\varepsilon \right) =\frac{2\pi \sqrt{%
2\varepsilon }}{\hbar ^{3}\sqrt{\det \mathbf{m}^{-1}}}\left\{ \frac{\exp
\left( i\sqrt{\Gamma _{0}^{2}\left( \mathbf{r}-\mathbf{r}^{\prime }\right) +%
\frac{8\varepsilon m_{zz}zz^{\prime }}{\hbar ^{2}}}\right) }{\sqrt{\Gamma
_{0}^{2}\left( \mathbf{r}-\mathbf{r}^{\prime }\right) +\frac{8\varepsilon
m_{zz}zz^{\prime }}{\hbar ^{2}}}}-\frac{\exp \left( i\Gamma _{0}\left(
\mathbf{r}-\mathbf{r}^{\prime }\right) \right) }{\Gamma _{0}\left( \mathbf{r}%
-\mathbf{r}^{\prime }\right) }\right\} .  \label{G_ellips}
\end{equation}%
Using Eqs.~(\ref{phi0_ell}) and (\ref{G_ellips}) we obtain the wave function
(\ref{psi0+g}) to first approximation in the strength of the impurity
potential. The modulus of this wave function is illustrated in Fig.~\ref%
{fig-ell-abs-phi} for a plane normal to the interface passing
through the impurity and the contact. The long axis of the
ellipsoid is rotated by $\pi /4$ around the $x-$axis, away from
the $y-$axis. An interference pattern is visible of partial waves
reflected from the impurity with those emanating from the contact
that is influenced by the anisotropy of the electronic band
structure.

The conductance in the limit of low temperatures, $T\rightarrow 0$, is
obtained from Eqs.~(\ref{flow_int}),(\ref{conduct}) by integration over all
directions of the momentum $\mathbf{p}$ and integration over the space
coordinate $\mathbf{\rho }$ in a plane $z=\mathrm{const}$ ($z>z_{0}$),
retaining only terms to first order in $g$ (i.e. ignoring multiple
scattering at the impurity site),
\begin{gather}
G^{\mathrm{ell}}\left( V\right) =G_{0}^{\mathrm{ell}}\left\{ 1-\frac{12\pi
^{2}gz_{0}^{2}\left( 2\varepsilon _{\mathrm{F}}\right) ^{3/2}}{\hbar ^{5}%
\sqrt{m_{zz}}\det \left[ \mathbf{m}^{-1}\right] }\frac{1}{\Gamma
_{0}^{4}\left( \varepsilon _{\mathrm{F}},\mathbf{r}_{0}\right) }\times
\right.  \label{G(V)} \\
\left. \left[ \left( 1-\frac{1}{\Gamma _{0}^{2}\left( \varepsilon _{\mathrm{F%
}},\mathbf{r}_{0}\right) }\right) \sin 2\Gamma _{0}\left( \varepsilon _{%
\mathrm{F}}+eV,\mathbf{r}_{0}\right) +\frac{1}{\Gamma _{0}\left( \varepsilon
_{\mathrm{F}},\mathbf{r}_{0}\right) }\cos 2\Gamma _{0}\left( \varepsilon _{%
\mathrm{F}}+eV,\mathbf{r}_{0}\right) \right] \right\} .  \notag
\end{gather}%
with $\Gamma _{0}\left( \varepsilon ,\mathbf{r}\right) $ given by (\ref%
{Ga_el}).

The amplitude of the conductance oscillations is maximal when $\Gamma _{0}$
is minimal. For a fixed depth $z_{0}$ this minimum occurs when the defect
position $\mathbf{\rho }_{0}$ is in the point $\mathbf{\rho }_{00}$ with
respect to the point contact at $\mathbf{r}=0$, where
\begin{equation}
\mathbf{\rho }_{00}=z_{0}\left(
\begin{array}{c}
m_{zz}/m_{zx} \\
m_{zz}/m_{zy}%
\end{array}%
\right) .  \label{x0y0}
\end{equation}%
The minimal value of the phase then becomes,
\begin{equation}
\Gamma _{00}=\Gamma_{0} \left( \varepsilon _{\mathrm{F}}+eV,\mathbf{\rho }%
_{00},z_{0}\right) =\frac{1}{\hbar }z_{0}\sqrt{2\left( \varepsilon _{\mathrm{%
F}}+eV\right) m_{zz}};  \label{Ga00_ell}
\end{equation}%
The phase $\Gamma _{00}$ corresponds to the extremal value of the chord of
the FS in the direction normal to the interface. $G_{0}^{\mathrm{ell}}$ in
Eq.~(\ref{G(V)}) is the conductance in the absence of a defect $\left(
g=0\right) $:
\begin{equation}
G_{0}^{\mathrm{ell}}=\frac{2e^{2}R^{4}\varepsilon _{\mathrm{F}}^{3}}{9\pi
\hbar ^{3}U^{2}\sqrt{\det \left[ \mathbf{m}^{-1}\right] }\sqrt{m_{zz}}}.
\label{G_c}
\end{equation}

\begin{figure}
\includegraphics[width=12cm,angle=0]{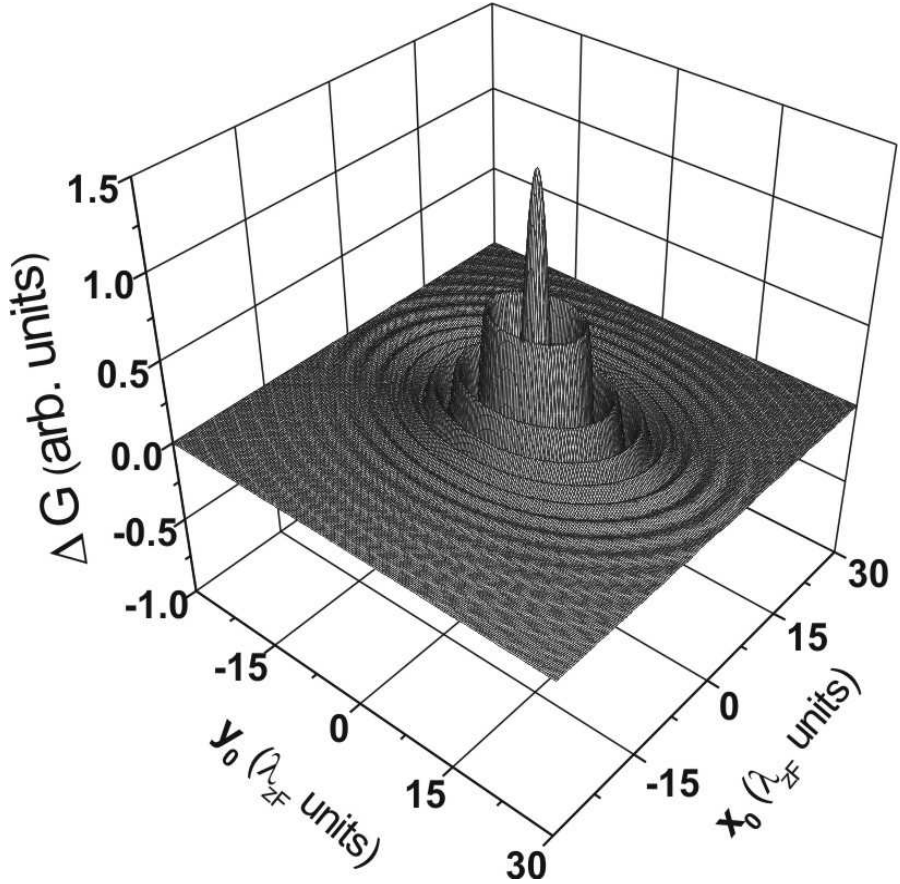}
\caption{Dependence of the oscillatory part of the conductance, $\Delta G$,
as a function of the position of the defect $\mathbf{\protect\rho }_{0}$ in
the plane $z=z_{0}$ for the same shape and orientation of the ellipsoidal FS
as in Fig.~\protect\ref{fig-ell-abs-phi}. The coordinates are measured in
units $\lambda _{z\mathrm{F}}=\hbar /\protect\sqrt{2m_{\mathrm{zz}}\protect\varepsilon _{\mathrm{F}}%
}$ and the defect sits at $z_{0}=5$. } \label{fig-ell-deltaG}
\end{figure}

Fig.~\ref{fig-ell-deltaG} shows a plot of the normalized conductance, $G^{%
\mathrm{ell}}\left( 0\right) /G_{0}^{\mathrm{ell}}$, Eq.~(\ref{G(V)}), for
the contact as a function of the position of the defect, $\mathbf{\rho }_{0}$%
, in the limit of low voltage, $V\rightarrow 0$. We find that $G^{\mathrm{ell%
}}$ is an oscillatory function of the defect position that reflects the
anisotropy of the FS and the oscillations are largest when the defect is
placed in the position $\mathbf{\rho }_{00}$, defined by Eq.~(\ref{x0y0}).

\begin{figure}
\includegraphics[width=8cm,angle=0]{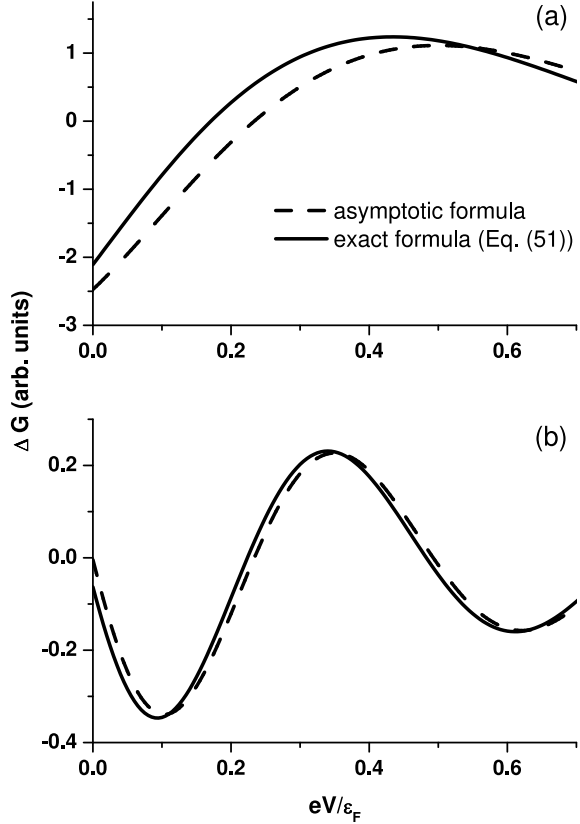}
\caption{Comparison of the oscillating part of the conductance for
an ellipsoidal FS calculated in the point $\mathbf{\rho }_{0}$%
 (\ref{x0y0}) of the maximum amplitude by using the asymptotic
(dashed curve) and exact (solid curve) formulas. The depths of the
defect are $z_{0}=4$ (a) and $z_{0}=10$ (b) in units of $\lambda
_{z\mathrm{F}}$. } \label{fig-comparision}
\end{figure}

For the ellipsoidal model FS the wave function and conductance have been
obtained exactly, within the framework of the model. For large distances $r$
and $r_{0}$ they transform into the asymptotic expressions, Eqs.~(\ref%
{psi_asym}), (\ref{G_asym}). We do not present the asymptotic form
explicitly but it agrees to within a term proportional to $\Gamma
_{0}^{-4}$ to the exact from, Eq.(\ref{G(V)}). In
Fig.~\ref{fig-comparision} we compare the results for the
calculations of the conductance by using the exact (\ref{G(V)})
and asymptotic (\ref{G_asym})\ expressions. The figure confirms
that for relatively small distances
(Fig.~\ref{fig-comparision}(a)) the asymptotic formula still
qualitatively describes the conductance very well and that for
larger distances (Fig.~\ref{fig-comparision}(b)) the two results
are in a good agreement. The parameters for the FS in
Figs.~\ref{fig-ell-deltaG} and \ref{fig-comparision} are the same
as those for Fig.~\ref{fig-ell-abs-phi}.

\section{Open Fermi surface}

\begin{figure}[tbp]
\includegraphics[width=10cm,angle=0]{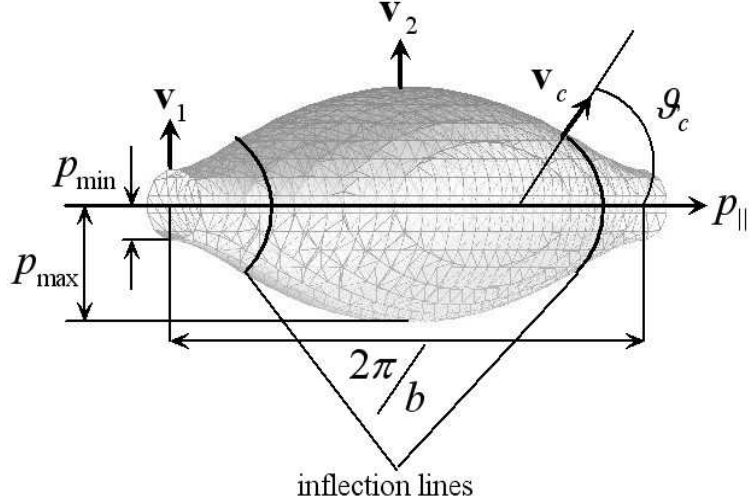}
\caption{Model of an open FS.}
\label{fig-openFS}
\end{figure}

The second model FS we want to discuss has the form of a corrugated cylinder
(Fig.~\ref{fig-openFS}), which is open along the direction $%
p_{\shortparallel }$,
\begin{equation}
\varepsilon \left( \mathbf{p}\right) =\frac{p_{\perp }^{2}}{2m}+\varepsilon
_{1}\sin ^{2}\frac{p_{\shortparallel }b}{2};\quad -\frac{\pi }{b}\leq
p_{\shortparallel }\leq \frac{\pi }{b};\quad \varepsilon >\varepsilon _{1},
\label{e_open1}
\end{equation}%
where $2\pi /b$ is the size of the Brillouin zone, and $m$ is an
effective mass. We further impose that the momentum perpendicular
to the symmetry axis of the FS remains finite,
\begin{equation}
p_{\bot \max }\leq p_{\bot }\leq p_{\bot \min };
\end{equation}%
where%
\begin{equation}
p_{\bot \max }\left( \varepsilon \right) =\sqrt{2m\varepsilon };\qquad
p_{\bot \min }\left( \varepsilon \right) =\sqrt{2m\left( \varepsilon
-\varepsilon _{1}\right) }  \label{p(e)}
\end{equation}%
are the maximal and minimal radii of the cylindrical surface, respectively.
As a consequence of rotational symmetry the Gaussian curvature $K$ (\ref%
{Gaus0}) of the surface depends only on $p_{\bot }$. The central part of the
surface (`belly') has a positive curvature $K>0$ while the ends near the
Brillouin zone boundary (`necks') have negative curvature. In the direction
perpendicular to the symmetry axis there are two partial waves propagating
with different parallel velocities, $\mathbf{v}_{1}$ and $\mathbf{v}_{2}$,
belonging to the parts of FS having opposite sign of $K$. Rotating away from
the perpendicular direction towards the axis the two solutions persist but
the two corresponding points on the FS move closer together until they merge
at the curve defined by $K=0$, the inflection line. For directions beyond
this angle (i.e. for $\theta <\theta _{c}$ in Fig.~\ref{fig-openFS}) no
propagating wave solutions exist. On the inflection line a unique solution
with velocity $\mathbf{v}_{c}$ is found.
For $\theta >\theta _{c}$ there are two stationary phase points $p_{\bot s}$
$\left( s=1,2\right) $ that satisfy Eq.~(\ref{st_point2}), corresponding two
different velocities $\mathbf{v}^{\left( +\right) }\left( p_{\bot s}\right) $
directed along the radius vector $\mathbf{r}$. The larger value, $p_{\bot 2}$%
, belongs to the belly of the FS $\left( K>0\right) $ and the smaller one, $%
p_{\bot 1}$, belongs to the neck $\left( K<0\right) $. At the inflection
line of the surface we have,
\begin{equation}
\left. \frac{\partial ^{2}p_{\shortparallel }\left( \varepsilon ,p_{\bot
}\right) }{\partial p_{\bot }^{2}}\right\vert _{p_{\bot }=p_{\bot 0}}=0,
\end{equation}%
which defines the value of perpendicular momentum $p_{\bot 0}$. From this
condition we obtain
\begin{equation}
p_{\bot 0}=\sqrt{p_{\bot \max }p_{\bot \min }}.  \label{p1}
\end{equation}%
The cone inside of which no propagating states exist is defined by the
condition
\begin{equation}
\left. \frac{v_{\shortparallel }}{v_{\bot }}\right\vert _{p_{\bot }=p_{\bot
0}}=\cot \vartheta _{c}=\frac{b}{2}\left( p_{\bot \max }-p_{\bot \min
}\right) \text{sgn}\left( v_{\shortparallel }\right) ,  \label{X}
\end{equation}%
where the components of the velocity $v_{\shortparallel }$ and $v_{\bot }$
are given by,
\begin{equation}
v_{\shortparallel }=\frac{\varepsilon _{1}b}{2}\sin \left( p_{\shortparallel
}b\right) ;\qquad v_{\bot }=\frac{p_{\bot }}{m}.
\end{equation}

In spite of the simplicity of the model FS (\ref{e_open1}), the
integrals in Eqs.~(\ref{psi0+g}),(\ref{conduct}) cannot be
evaluated analytically. We can only discuss the asymptotic
behavior for $r_{0}\gg \lambda _{\mathrm{F}}$.
Qualitatively this result should also be valid for $r_{0}>\lambda _{\mathrm{F%
}}$. For the directions that have two stationary phase points,
having opposite signs of the Gaussian curvature,
Eq.~(\ref{G_asym}) acquires the form
\begin{eqnarray}
G^{\mathrm{op}} &=&G_{0}^{\mathrm{op}}\left( 1+\frac{g\cos^{2}\vartheta(%
\mathbf{n}_{0})}{\left( 2\pi \hbar \right) ^{3}\hbar \left\langle \left(
v_{z}^{\left( +\right) }\right) ^{2}\right\rangle \nu \left( \varepsilon _{%
\mathrm{F}}\right) r_{0}^{2}}\sum\limits_{s,s^{\prime }=1,2}\frac{1}{\sqrt{%
\left\vert K_{0}^{\left( s\right) }\left( \varepsilon _{\mathrm{F}},\mathbf{n%
}_{0}\right) K_{0}^{\left( s^{\prime }\right) }\left( \varepsilon _{\mathrm{F%
}},\mathbf{n}_{0}\right) \right\vert }}\times \right.  \notag  \label{G_op}
\\
&&\left. \cos \left( \Gamma _{0}^{\left( s\right) }\left( \varepsilon _{%
\mathrm{F}}+eV,\mathbf{r}_{0}\right) +\frac{\pi }{2}\left( 1-s\right)
\right) \sin \left( \Gamma _{0}^{\left( s^{\prime }\right) }\left(
\varepsilon _{\mathrm{F}}+eV,\mathbf{r}_{0}\right) +\frac{\pi }{2}\left(
1-s^{\prime }\right) \right) \right)
\end{eqnarray}
where $G_{0}^{\mathrm{op}}$ is the conductance of the contact
without defect given by  Eq.~(\ref
{G0_arbit}), $\Gamma _{0}^{\left( s\right)}(\varepsilon ,%
\mathbf{r}) = \Gamma (\mathbf{p}_{t,s}^{\left(\mathrm{st}\right)},\mathbf{r}%
) $, and $K_{0}^{\left( s\right) }(\varepsilon ,\mathbf{n}) = K(\mathbf{p}%
_{t,s}^{\left( \mathrm{st}\right) },\varepsilon)$.

The appearance of the conductance oscillations depends strongly on the
orientation of the FS with respect to the interface. Below we will consider
two specific orientations, having the axis of the FS either perpendicular or
parallel to the interface.

\subsection{Direction of open FS perpendicular to the interface}

When the iso-energy surface is open along the contact axis $z$ the
components of the momenta in Eq.~(\ref{e_open1}) are $p_{\perp }=\sqrt{%
p_{x}^{2}+p_{y}^{2}}$ and $p_{\shortparallel }=p_{z}$. In this case the
conductance of the clean contact (without defect) becomes,
\begin{equation}
G_{0}^{\mathrm{op}}=\frac{\pi e^{2}R^{4}m^{2}\varepsilon _{1}^{4}b^{2}}{%
8\hbar ^{3}U^{2}}.
\end{equation}%
From Eq.~(\ref{st_point2}) the stationary phase points for the iso-energy
surface are,
\begin{eqnarray}
p_{\bot s}^{2} &=&\frac{1}{2}\left( p_{\bot \max }^{2}+p_{\bot \min }^{2}-%
\frac{4}{b^{2}}\cot ^{2}\vartheta +\right.  \label{p012} \\
&&\left. \left( -1\right) ^{s}\sqrt{\left( p_{\bot \max }^{2}+p_{\bot \min
}^{2}-\frac{4}{b^{2}}\cot ^{2}\vartheta \right) ^{2}-4p_{\bot \max
}^{2}p_{\bot \min }^{2}}\quad \right) .  \notag
\end{eqnarray}%
The angle $\vartheta =\arccos (z/r)$ is defined by the direction of the
radius vector $\mathbf{r}$. The Gaussian curvature $K_{0}$ and the phase $%
\Gamma _{0}$ in the points (\ref{p012}) are given by relations,
\begin{equation}
K_{0}^{\left( s\right) }\left( \varepsilon ,\mathbf{n}\right) =\frac{%
b^{2}\left( -1\right) ^{s}\sin ^{4}\vartheta }{4p_{\bot s}^{2}}\sqrt{\left(
p_{\bot \max }^{2}+p_{\bot \min }^{2}-\frac{4}{b^{2}}\cot ^{2}\vartheta
\right) ^{2}-4p_{\bot \max }^{2}p_{\bot \min }^{2}}\quad ;  \label{K0_open1}
\end{equation}%
\begin{equation}
\Gamma _{0}^{\left( s\right) }\left( \varepsilon ,\mathbf{r}\right) =\frac{1%
}{\hbar }p_{\bot s}\sqrt{x^{2}+y^{2}}+\frac{2z}{\hbar b}\arcsin \sqrt{\frac{%
2m\varepsilon -p_{\bot s}^{2}}{2m\varepsilon _{1}}}\quad .  \label{Ga0_open1}
\end{equation}%
The angle $\vartheta $ in Eqs.~(\ref{p012}),(\ref{K0_open1}) is contained
within the interval $\vartheta _{c}\leq \vartheta \leq \pi /2,$ where the $%
\vartheta _{c}$ is given by Eq.~(\ref{X}).

\begin{figure}[tbp]
\includegraphics[width=10cm,angle=0]{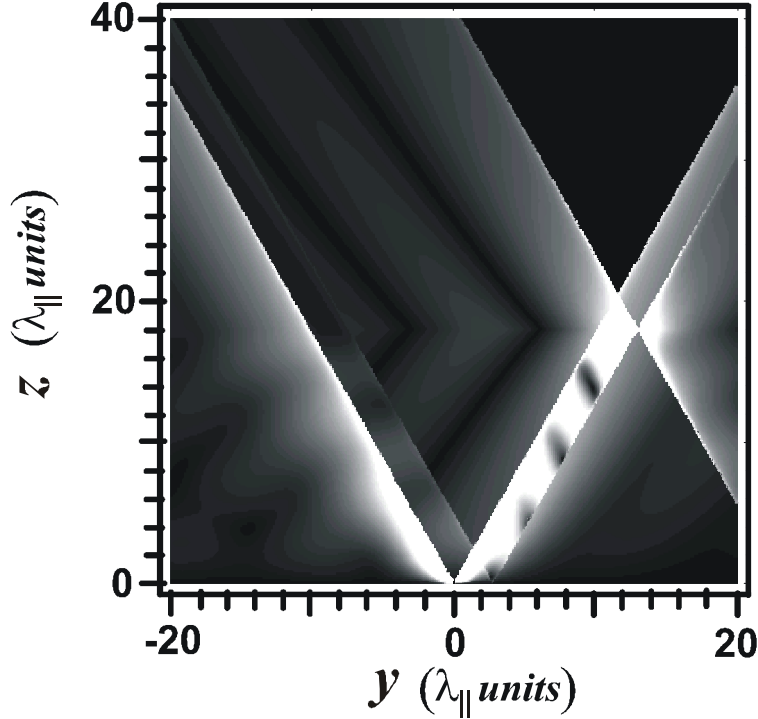}
\caption{Gray-scale plot of the modulus of the wave function in
the plane $x=0$ for a warped cylindrical FS having the open
direction along the contact axis $z$. The
coordinates are measured in units of $\lambda _{||}=\hbar /\protect\sqrt{2m\protect%
\varepsilon}$. The parameters used in the model are $\protect\varepsilon %
_{1}/\protect\varepsilon=0.9$, $b\protect\sqrt{2m\protect\varepsilon}=5.2$%
, and a defect sits at $\mathbf{r}_{0}=\left( 0,13,18\right) .$}
\label{fig-modulus-openFS}
\end{figure}

The modulus of the wave function is plotted in
Fig.~\ref{fig-modulus-openFS}. For the calculation of the wave
function, Eq.~(\ref{psi_asym}), we used formulas (\ref{K0_open1})
for the curvature and (\ref{Ga0_open1}) for the phase in the
asymptotic expression for the integral $\Lambda ^{\mathrm{as}}$
(\ref{lam}). Although, strictly speaking Eq.~(\ref{psi_asym}) is
not applicable in the vicinity of the contact and near the defect,
nor inside the classically inaccessible region,
Fig.~\ref{fig-modulus-openFS} illustrates the main features of
this problem. One observes the interference of the two partial
waves with different velocities, the existence of a forbidden
cone, the anisotropy of the waves scattered by the defect, and the
enhanced wave function amplitude near the edge of the forbidden
cone.

At the inflection lines, where $K=0$ and $\theta $ is given by
Eq.~(\ref{X}), the square root in Eq.~(\ref{p012}) is equal to
zero. For this direction two stationary phase points merge and the
electron velocity is directed along the cone of the classically
forbidden region. The asymptotic expression for the conductance
(\ref{G_op}) diverges at these points, which implies that the
third derivative of the phase, Eq.~(\ref{phase}),
with respect to $p_{\bot }$ must be taken into account. When the vector $%
\mathbf{r}_{0}$ connecting the point contact to the defect lies along the
cone of the forbidden region the conductance oscillations have maximal
amplitude and the conductance takes the form,
\begin{equation*}
G_{\max }^{\mathrm{op}}=G_{0}^{\mathrm{op}}\left( 1-\frac{Cg}{m\varepsilon
_{1}}\left( \frac{b^{2}}{\hbar ^{4}z_{0}^{5}}\right) ^{1/3}\sqrt{p_{\bot
\max }p_{\bot \min }}\left( p_{\bot \max }-p_{\bot \min }\right) ^{3}\sin
\left( 2\Gamma _{00}-\frac{5\pi }{6}\right) \right) ,
\end{equation*}%
where%
\begin{equation}
\Gamma _{00}=\Gamma _{0}\left( p_{\bot 0},z_{0}\right)
=\frac{2z_{0}}{b\hbar }\left. \left( \frac{\sqrt{p_{\bot \max
}p_{\bot \min }}}{p_{\bot \max }-p_{\bot \min }}+\arcsin
\sqrt{\frac{p_{\bot \max }^{2}-p_{\bot \max }p_{\bot \min
}}{2m\varepsilon _{1}}}\right) \right\vert _{\varepsilon
=\varepsilon _{\mathrm{F}}+eV},  \label{Ga00}
\end{equation}%
and $C$ is a numerical constant, $C\simeq 1.97\cdot 10^{-4}.$ The
energy dependencies of $p_{\bot \max }$ and $p_{\bot \min }$ are
given by Eq.~(\ref{p(e)}).

\begin{figure}[tbp]
\includegraphics[width=10cm,angle=0]{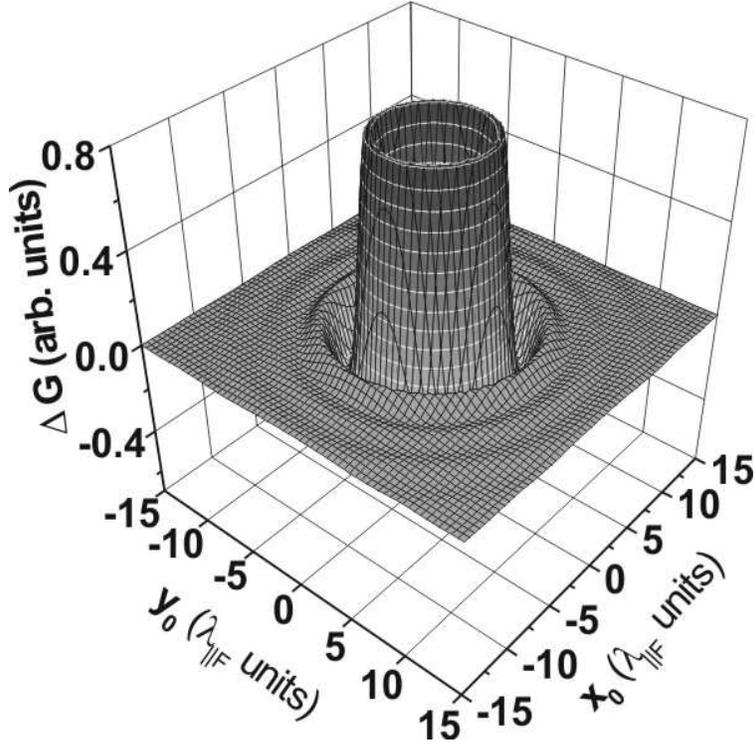}
\caption{Dependence of the oscillatory part $\Delta G$ of the conductance,
as a function of the lateral position of the defect $\mathbf{\protect\rho }%
_{0}$ in the plane $z=z_{0}.$ The open direction of the FS is oriented
perpendicular to the interface. The coordinates are measured in units of $%
\lambda _{||\mathrm{F}}=\hbar
/\protect\sqrt{2m\protect\varepsilon_{\mathrm{F}}}$. The
parameters used for the
model FS (\protect\ref{e_open1}) are $\protect\varepsilon _{1}/\protect%
\varepsilon_{\mathrm{F}}=0.9$,
$b\protect\sqrt{2m\protect\varepsilon_{\mathrm{F}}}=6$, and a
defect sits at a depth of $z_{0}=11.$}
\label{fig-deltaG-open-perp}
\end{figure}

Fig.~\ref{fig-deltaG-open-perp} shows a plot of the oscillatory part $\Delta
G$ of the conductance $G^{\mathrm{op}}\left( 0\right) $, Eq.~(\ref{G_op}),
as a function of the lateral position of the defect $\mathbf{\rho }_{0}$ for
a fixed distance $z_{0}$ from the interface.
The oscillation pattern has a `dead' region in the center, corresponding to
defect positions inside the classically inaccessible part of the metal for
electrons injected by the point contact. This region is defined by the cone (%
\ref{X}) and its radius $\rho _{00}=z_{0}/\cot \vartheta _{c}$ depends on
the depth of the defect under surface. The oscillations in the conductance
are largest when the defect is placed at the edge of the cone $\rho
_{0}=\rho _{00}$. In this case the defect is positioned in a direction of
velocity belonging to the inflection line of the FS and the electron flux in
this direction is maximal.

\subsection{Direction of open FS parallel to the interface}

The second orientation we want to discuss is that with the FS (\ref{e_open1}%
) having its open direction (the axis) parallel to interface, with $\mathbf{p%
}_{\perp }=\left( p_{x},p_{z}\right) $ and $p_{\shortparallel }=p_{y}$. The
existence of a classically inaccessible region for this geometry leads to a
strongly anisotropic current density in the $xy$ - plane. The expression for
the point contact conductance without defect $G_{0}^{\mathrm{op}}$ becomes,
\begin{equation}
G_{0}^{\mathrm{op}}=\frac{e^{2}R^{4}\pi \left( 2\varepsilon -\varepsilon
_{1}\right) ^{2}}{4\hbar ^{3}b^{2}U^{2}}.
\end{equation}%
The expressions for the phase, Eq.~(\ref{Ga_el}), and the Gaussian
curvature, Eq.~(\ref{Gaus0}), now read,
\begin{equation}
\Gamma _{0}^{\left( s\right) }\left( \varepsilon ,\mathbf{r}\right) =\frac{1%
}{\hbar }\left( \sqrt{\left( x^{2}+z^{2}\right) \left( 2m\left( \varepsilon
-\varepsilon _{1}\lambda _{s}\right) \right) }+\frac{2\left\vert
y\right\vert }{b}\arcsin \sqrt{\lambda _{s}}\right) ,  \label{phase3}
\end{equation}%
and
\begin{equation}
K_{0}^{\left( s\right) }\left( \varepsilon ,\mathbf{n}\right) =\left(
-1\right) ^{s}\frac{\varepsilon _{1}b^{2}\sin ^{4}\theta }{4\left(
\varepsilon -\varepsilon _{1}\lambda _{s}\right) }\sqrt{\left( 1+\frac{2\cot
^{2}\theta }{m\varepsilon _{1}b^{2}}\right) ^{2}-\frac{8\varepsilon \cot
^{2}\theta }{m\varepsilon _{1}^{2}b^{2}}},  \label{Gaus2}
\end{equation}%
respectively. For this geometry we use spherical coordinates, with
$\theta $ the angle between the vector $\mathbf{r}$ and the
$y$-axis. The variables $ \lambda _{1,2}$ have been obtained from
Eq.~(\ref{st_point2}),
\begin{equation}
\lambda _{s}=\frac{1}{2}\left\{ 1+\frac{2\cot ^{2}\theta }{m\varepsilon
_{1}b^{2}}+\left( -1\right) ^{s}\sqrt{\left( 1+\frac{2\cot ^{2}\theta }{%
m\varepsilon _{1}b^{2}}\right) ^{2}-\frac{8\varepsilon \cot ^{2}\theta }{%
m\varepsilon _{1}^{2}b^{2}}}\right\} .  \label{lam12}
\end{equation}%
The first stationary phase point, $\lambda _{1}$, corresponds to a positive
Gaussian curvature and the second one, $\lambda _{2}$, to a negative
curvature. For $\theta =\vartheta _{c}$, Eq.(\ref{X}), they become equal,
\begin{equation}
\lambda _{1}=\lambda _{2}=\frac{1}{2m\varepsilon _{1}}\left( p_{\bot \max
}^{2}-\sqrt{p_{\bot \max }^{2}p_{\bot \min }^{2}}\right) ,
\end{equation}%
and the curvature (\ref{Gaus2}) vanishes.

\begin{figure}
\includegraphics[width=10cm,angle=0]{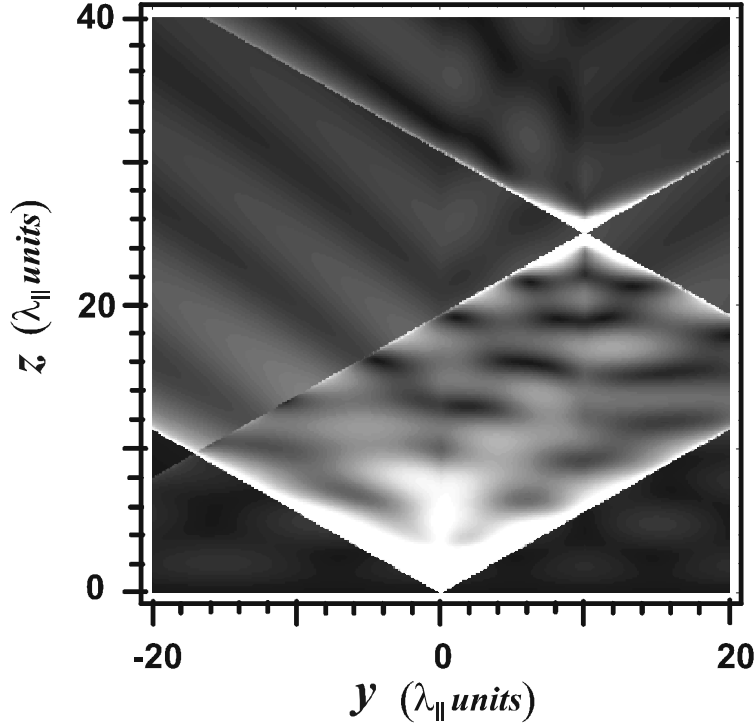}
\caption{Gray-scale plot of the modulus of the wave function in
the plane $x=0$ for the warped cylindrical Fermi surface with the
open direction along the $y$-axis and parallel to the plane of
interface. The coordinates are measured in units of $\lambda
_{||}$. The Fermi surface parameters are $\varepsilon
_{1}/\varepsilon =0.9$, $b\sqrt{2m\varepsilon }=5.2$, and the
defect position is $\mathbf{r}_{0} =\left( 0,10,25\right)$.}
\label{fig-mod-openFS-par}
\end{figure}

Figure \ref{fig-mod-openFS-par}, acquired by using Eqs.
(\ref{psi_asym}), (\ref{phase3}) and (\ref{Gaus2}), illustrates
the interference of waves with different velocities the electrons
emerging from the contact and the interference with the waves
scattered by the defect. For this geometry the classically
inaccessible region is found near the interface to both sides of
the contact.

\begin{figure}
\includegraphics[width=10cm,angle=0]{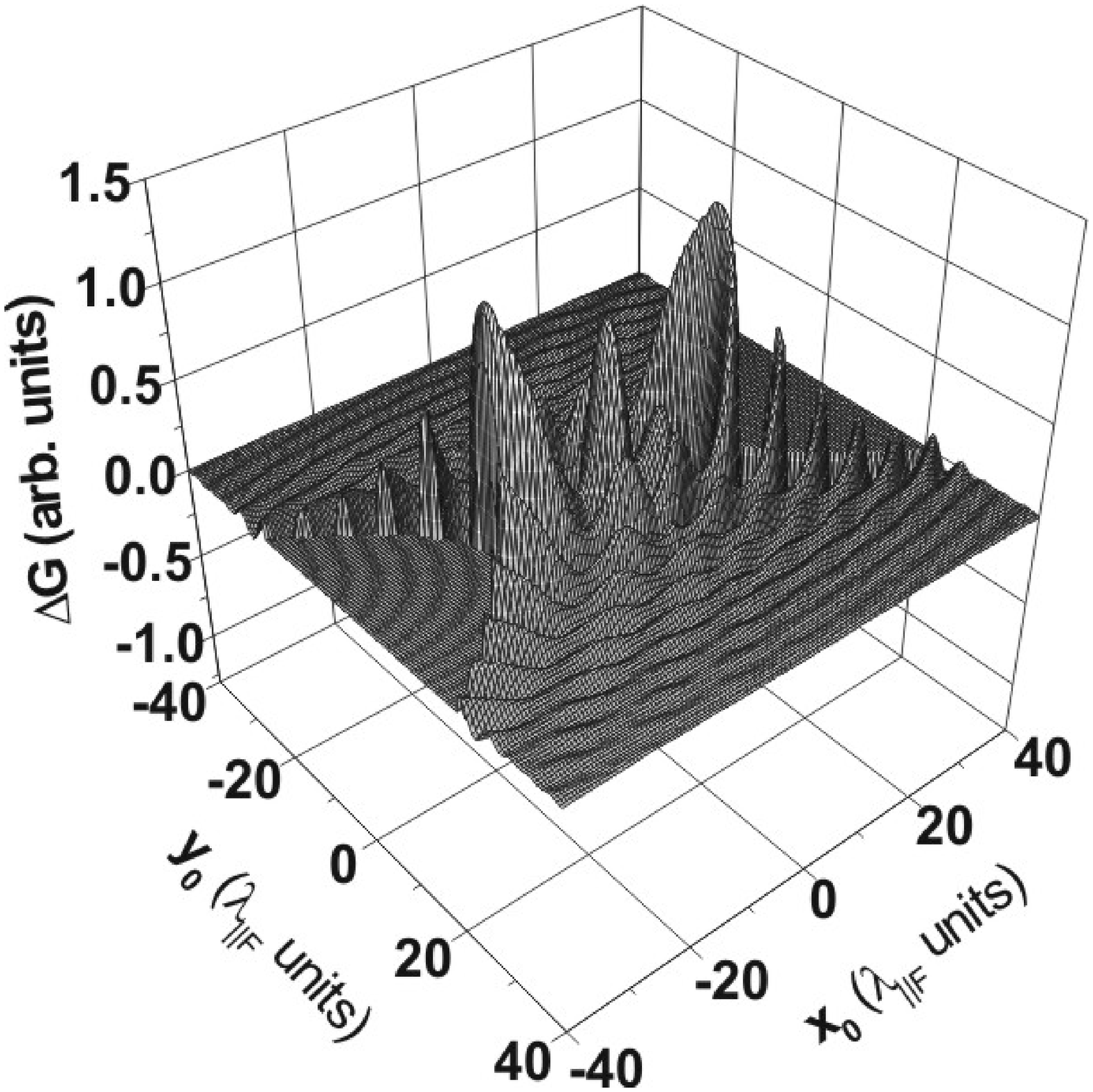}
\caption{Dependence of the oscillatory part $\Delta G$ of the conductance,
as a function of the lateral position of the defect $\mathbf{\protect\rho }%
_{0}$ in the plane $z=z_{0}$. The open direction of the FS is
oriented parallel to the interface along the $y-$direction. The
coordinates are measured in units of $\lambda _{||\mathrm{F}}$.
The parameters used for the model FS (\protect\ref{e_open1}) are $\protect%
\varepsilon _{1}/\protect\varepsilon_{\mathrm{F}} =0.9$, $b\protect\sqrt{2m\protect%
\varepsilon_{\mathrm{F}}}=6$, and a defect sits at a depth of
$z_{0}=11$. } \label{fig-conductance-openFS-par}
\end{figure}

Figure~\ref{fig-conductance-openFS-par} shows a plot of the
oscillatory part $\Delta G$ of the conductance $G^{\mathrm{op}}(0)
$ as a function of the lateral position of the defect
$\mathbf{\rho }_{0}$ for a fixed distance $z_{0}$ from the
interface. In this case two `dead' regions appear symmetrically
with respect to the center of the oscillation pattern along the
open direction of the FS. The center of the pattern corresponds to
a defect sitting on the axis of the
contact, $\mathbf{\rho }_{0}=0$, for which $\lambda _{1}=0$, $\lambda _{2}=1$%
, $\sin \theta =1$, and $\cos \varphi =1$. At this point Eq.~(\ref{G_asym})
takes the form
\begin{eqnarray}
G^{\mathrm{op}}\left( V\right) &=&G_{0}^{\mathrm{op}}\left\{ 1+\frac{4g}{%
z_{0}^{2}\pi ^{2}\hbar \left( p_{\bot \max }^{2}+p_{\bot \min }^{2}\right)
\varepsilon _{1}b}\left[ p_{\bot \min }^{2}\sin \left( \frac{2p_{\bot \min
}(\varepsilon _{F}+eV)z_{0}}{\hbar }\right) \right. \right.  \label{Gopen2}
\\
&&-p_{\bot \max }^{2}\sin \left( \frac{2p_{\bot \max }(\varepsilon
_{F}+eV)z_{0}}{\hbar }\right)  \notag \\
&&\left. \left. +2p_{\bot \max }p_{\bot \min }\cos \left( \frac{p_{\bot \max
}(\varepsilon _{F}+eV)z_{0}+p_{\bot \min }(\varepsilon _{F}+eV)z_{0}}{\hbar }%
\right) \right] \right\} .  \notag
\end{eqnarray}

\section{\protect\bigskip Discussion}

We have analyzed the oscillatory voltage dependence of the
conductance of a tunnel junction in the presence of an elastic
scattering center located inside the bulk for metals with an
anisotropic FS. These oscillations result from electron waves
being scattered by the defect and reflected back by the contact,
interfering with electrons that are directly transmitted through
the contact. The introduction of anisotropic electron movement
beams the following implication: several points on the FS may
share the same direction
of the group velocity vector $\mathbf{v}$ whereas other directions for $%
\mathbf{v}$ can be absent. Two non-spherical shapes for the FS have been
investigated: the ellipsoid and the corrugated cylinder (open surface).

Contrary to the case of a spherical FS \cite{Avotina1} in the ellipsoidal
model (\ref{energy}) the center of the conductance oscillation pattern does
not need to coincide with the actual position of the defect but is displaced
over a vector $\mathbf{\rho }_{00}=z_{0}\left(
m_{zz}/m_{zx},m_{zz}/m_{zy}\right) $. When the STM tip is placed at this
point the oscillatory part of the conductance is given by,
\begin{equation}
\Delta G(V)\propto \cos \left( \frac{2}{\hbar }z_{0}\sqrt{2\left(
\varepsilon _{F}+eV\right) m_{zz}}\right) .\quad  \label{G_el_osc}
\end{equation}%
The oscillation period depends on $1/m_{zz}$, the component of the
tensor of inversive mass (\ref{m^(-1)}) for motion in the
$z$-direction, and on the depth $z_{0}$ of the defect. This allows
us in principle to map out the positions of defects, as long as
the shape of the FS is known. Apart from the period of the
oscillations, there is also information in the amplitude. Since
short periods should correspond to small amplitudes this may be
used for a test of consistency. However, quantitatively the
amplitude is also influenced by unknown factors such as the defect
scattering efficiency. The ellipsoidal FS is exceptional in that
the problem can be solved exactly. This allows us to compare the
calculation with the asymptotic approximation, and this shows that
the approximation works very well for distances larger than
$\lambda _{F}$.

In the case of the corrugated cylinder (\ref{e_open1}) the open
necks cause cones with opening angle $2\vartheta _{c}$ (defined by
the inflection line of the FS) to be classically inaccessible. If
the orientation is such that the open direction is orthogonal to
the surface this will result in a `dead' region with radius
$z_{0}/\cot \vartheta _{c}$ (\ref{X}) where no conductance
fluctuations can be observed. Thus, by measuring the size of this
dead region we directly obtain the position of the defect.  The
oscillation amplitude will be maximal at the border of the dead
region, since the current density will be highest in the direction
of the group velocity at the inflection line. In analogy to a
hurricane the `eye' is surrounded by a ring of intense currents.
Such rings of high amplitude oscillations have already been
reported very recently in experiments on Ag and Cu(111) surfaces
\cite{Wenderoth}. For our model FS, along this border the
oscillating part of the conductance is, apart from a phase factor,
described by,
\begin{equation}
\Delta G(V)\propto \sin \left( \frac{4z_{0}}{b\hbar }\left. \left( \frac{%
\sqrt{p_{\bot \max }p_{\bot \min }}}{p_{\bot \max }-p_{\bot \min }}+\arcsin
\sqrt{\frac{p_{\bot \max }^{2}-p_{\bot \max }p_{\bot \min }}{%
2m\varepsilon_{1}}}\right) \right\vert _{\varepsilon =\varepsilon
_{F}+eV}\right) ,
\end{equation}
where $p_{\bot \max }$ and $p_{\bot \min }$ are the maximal and minimal
radii of the surface of constant energy in the direction perpendicular to
the axis of the cylinder (\ref{p(e)}), $\varepsilon_{1}$ is the amplitude of
corrugation of the FS, and $2\pi/b$ is the size of the Brillouin zone. Again
we find that the depth of the defect is determining the oscillation period,
so that for given FS parameters this information can be exploited to
investigate the structure of the metal below the surface.

If the open direction is parallel with the surface the highest amplitude
will be found with the STM tip straight above the defect. For $\mathbf{\rho }%
_{0}=0$ the conductance oscillations are described by Eq.~(\ref{Gopen2}).
Clearly, the oscillation pattern is more complicated than that from Eq.~\ref%
{G_el_osc}, since there are contributions from the belly as well as from the
neck parts of the FS, plus a sum frequency. For small necks the signal will
be dominated by the oscillation due to the belly.

Although the two models presented in this paper are still rather
artificial, they provide insights that are quite valuable for
experimental work in this field. The most prominent conclusion is
that the regular oscillations due to convex parts of the FS, that
behave as for the isotropic FS discussed previously
\cite{Avotina1}, will often be dominated by signals due to special
directions. On any surface that features regions of zero
curvature, the strongest conductance fluctuations will come from
electrons travelling with the group velocity of that region. Not
only does this hold for the inflection lines around the necks in
the (111) direction of e.g. Cu, Ag or Au, it is also true for the
almost flat facets in the (110) direction of the same metals. In
the case of an inflection line the signal will decay as
$r^{-5/3}$, whereas for a flat facet the signal does not decay at
all. This effect can be exploited for imaging defects up to much
larger depths than previously estimated. The particular shape of
the FS for nearly all metals contain many detailed features that
will allow us to check the validity of the conclusions drawn from
the measured data.

We thank M. Wenderoth for communicating his unpublished results.
Ye.S.A. is supported by a grant of the European INTAS Young
Scientists program (No 04-83-3750) and Yu.A.K. was supported by
the European Erasmus Mundus program on Nanoscience.


\begin{thebibliography}{99}
\bibitem{Ludoph1} B. Ludoph, M.H. Devoret, D. Esteve, C. Urbina and J.M. van
Ruitenbeek, Phys. Rev. Lett., \textbf{82,} 1530-3, (1999).

\bibitem{Untiedt} C. Untiedt, G. Rubio Bollinger, S. Vieira, and N. Agra{%
\"{\i}}t, Phys. Rev. B, \textbf{62}, 9962 (2000).

\bibitem{Ludoph} B. Ludoph and J. M. van Ruitenbeek, Phys. Rev. B, \textbf{61%
}, 2273 (2000).

\bibitem{Kempen} A. Halbritter, Sz. Csonka, G. Mih{\'{a}}ly, O. I.
Shklyarevskii, S. Speller, and H. van Kempen, Phys. Rev. B, \textbf{69},
121411 (2004).

\bibitem{Namir} A. Namiranian, Yu. A. Kolesnichenko, and A. N. Omelyanchouk,
Phys. Rev. B, \textbf{61}, 16796 (2000).

\bibitem{Avotina2} Ye. S. Avotina, and Yu. A. Kolesnichenko, Fiz. Nizk.
Temp., \textbf{30, } 209 (2004) [J. Low Temp. Phys., \textbf{30}, 153
(2004)].

\bibitem{Avotina3} Ye. S. Avotina, A. Namiranian, and Yu. A. Kolesnichenko,
Phys. Rev. B, \textbf{70}, 075908 (2004).

\bibitem{Avotina1} \bigskip Ye. S. Avotina, Yu. A. Kolesnichenko, A.N.
Omelyanchouk, A.F. Otte, and J.M. Ruitenbeek, Phys. Rev. B \textbf{71},
115430 (2005).

\bibitem{LAK} I.M. Lifshits, M.Ya. Azbel', and M.I. Kaganov, \ "Electron
theory of metals", New York, Colsultants Bureau (1973).

\bibitem{Kosevich} A.M. Kosevich, Fiz. Nizk. Temp., \textbf{11}, 1106 (1985)
[Sov. J. Low Temp. Phys., \textbf{11}, 611 (1985)].

\bibitem{Ustinov} V.V. Ustinov, D.Z. Khusainov, Phys. Stat. Sol., \textbf{%
B134, }313 (1986).

\bibitem{KMO} I. O. Kulik, Yu. N. Mitsai, and A. N. Omelyanchouk, Zh. Exp.
Teor. Fiz., \textbf{63}, 1051 (1974).

\bibitem{ISh} I. F. Itskovich and R. I. Shekhter, Fiz. Nizk. Temp., \textbf{%
11}, 373 (1985) [Sov. J. Low Temp. Phys., \textbf{11}, 202 (1985)].

\bibitem{Flugge} S. Fl\"{u}gge, Practical Quantum Mechanics, V.1,
Springer-Verlag (1971).

\bibitem{Korn} G. Korn, T.Korn, "Mathematical Handbook", McGraw-Hill Company
(1968).

\bibitem{Azbel} M. Ya. Azbel, Phys. Rev. B, \textbf{43}, 2435 (1991).

\bibitem{Fedoriuk} V. P. Maslov, M. V. Fedoriuk : Semi-classical
approximation in quantum mechanics, Dordrecht: D.Reidel Publishing Company
(1981).

\bibitem{Wenderoth} A. Weismann, M. Wenderoth, N. Quaas, and R,G. Ulbrich,
`Focusing of bulk electrons in noble metals on the atomic scale',
unpublished.
\end{thebibliography}
\end{document}